\documentclass[%
 reprint,
superscriptaddress,
 amsmath,amssymb,
 aps,
prb,
floatfix,
]{revtex4-1}

\usepackage{graphicx}
\usepackage{dcolumn}
\usepackage{bm}
\usepackage[colorlinks=true,allcolors=blue]{hyperref}
\hypersetup{breaklinks=true}

\usepackage{array,booktabs,tabularx}
\usepackage[caption=false]{subfig}
\usepackage[USenglish]{babel}
\usepackage{braket}

\newcommand{\beq}{\begin{equation}}
\newcommand{\eeq}{\end{equation}}
\newcommand{\beqn}{\begin{eqnarray}}
\newcommand{\eeqn}{\end{eqnarray}}

\begin{document}
\selectlanguage{USenglish}
\preprint{APS/123-QED}

\title{Anharmonicity and the isotope effect in superconducting lithium at high pressures: a first-principles approach}

\author{Miguel Borinaga}
\affiliation{Centro de F\'isica de Materiales CFM, CSIC-UPV/EHU, Paseo Manuel de
             Lardizabal 5, 20018 Donostia/San Sebasti\'an, Basque Country, Spain}
\affiliation{Donostia International Physics Center
             (DIPC), Manuel Lardizabal pasealekua 4, 20018 Donostia/San
             Sebasti\'an, Basque Country, Spain}
\author{Unai Aseginolaza}
\affiliation{Centro de F\'isica de Materiales CFM, CSIC-UPV/EHU, Paseo Manuel de
             Lardizabal 5, 20018 Donostia/San Sebasti\'an, Basque Country, Spain}
\affiliation{Donostia International Physics Center
             (DIPC), Manuel Lardizabal pasealekua 4, 20018 Donostia/San
             Sebasti\'an, Basque Country, Spain}
\author{Ion Errea}
\affiliation{Donostia International Physics Center
             (DIPC), Manuel Lardizabal pasealekua 4, 20018 Donostia/San
             Sebasti\'an, Basque Country, Spain}
\affiliation{Fisika Aplikatua 1 Saila, Bilboko Ingeniaritza Eskola,
             University of the Basque Country (UPV/EHU), Rafael Moreno ``Pitxitxi'' Pasealekua 3, 48013 Bilbao,
             Basque Country, Spain}
\author{Matteo Calandra}
\affiliation{IMPMC, UMR CNRS 7590, Sorbonne
Universit\'es - UPMC Univ. Paris 06, MNHN, IRD, 4 Place Jussieu,
F-75005 Paris, France}
\author{Francesco Mauri}
\affiliation{Dipartimento di Fisica, Universit\`a di Roma La Sapienza, Piazzale Aldo Moro 5, I-00185 Roma, Italy}            
\author{Aitor Bergara}
\affiliation{Centro de F\'isica de Materiales CFM, CSIC-UPV/EHU, Paseo Manuel de
             Lardizabal 5, 20018 Donostia/San Sebasti\'an, Basque Country, Spain}
\affiliation{Donostia International Physics Center
             (DIPC), Manuel Lardizabal pasealekua 4, 20018 Donostia/San
             Sebasti\'an, Basque Country, Spain}
\affiliation{Departamento de F\'isica de la Materia Condensada,  University of the Basque Country (UPV/EHU), 48080 Bilbao, 
             Basque Country, Spain}

\date{\today}

\begin{abstract}

Recent experiments \cite{Schaeffer2015} have shown that lithium presents an extremely anomalous isotope effect in the 15-25 GPa pressure range. In this article we have calculated the anharmonic phonon dispersion of 
$\mathrm{^7Li}$ and  $\mathrm{^6Li}$ under pressure, their superconducting transition temperatures, and the associated isotope effect. We have found a huge anharmonic renormalization of a transverse acoustic soft mode 
along $\Gamma$K in the fcc phase, the expected structure at the pressure range of interest. In fact, the anharmonic correction dynamically stabilizes the fcc phase above 25 GPa. However, we have not found any anomalous scaling of 
the superconducting temperature with the isotopic mass. Additionally, we have also analyzed whether the two lithium isotopes adopting different structures could explain the observed anomalous behavior. According to 
our enthalpy calculations including zero-point motion and anharmonicity it would not be possible in a stable regime.  

\end{abstract}

\maketitle


\section{Introduction}\label{introduction}

The strongly anomalous isotope effect recently measured in lithium in the 15-25 GPa pressure range\cite{Schaeffer2015} brought this element back under the spotlight. 
The reported superconducting critical temperatures ($\mathrm{\mathrm{T_c}}$) contrast starkly with the BCS theory, where $\mathrm{\mathrm{T_c}}$ is expected to scale as $\propto 1/M^{\alpha}$,
with $M$ being the atomic mass and $\alpha$ the isotope coefficient (0.5 within the BCS theory). Actually, for most phonon mediated superconductors, $\alpha$ does not deviate
much from 0.5. However, the above mentioned experiment shows a highly erratic behavior of $\alpha$ as a function of pressure, with values ranging from 1 to 4 from 15 to 21 GPa, decreasing sharply between 21 and 25 GPa, where it even becomes negative, with values as low as -2.

It is just another fascinating example of the rich and exotic phenomena emerging in lithium under pressure. The lightest metal on the periodic table shows a nearly free-electron bcc structure at 
ambient conditions\cite{Seitz1935}. Although it could be expected to evolve to an even more free-electron like system with increasing pressure, it has been shown that pressure 
not only induces several structural transformations~\cite{Guillaume2011,Hanfland2000,1742-6596-121-5-052003,Schaeffernaturecomms,Ackland2017}, but also gives rise to a plethora of fascinating physical properties\cite{springerlink:10.1140/epjb/e2011-10972-9}.
For instance, lithium becomes a semiconductor 
near 80 GPa\cite{matsuoka:186}, it shows a maximum in the melting line \cite{PhysRevLett.104.185701} and melts below ambient temperature (190 K) at around 50 GPa\cite{Guillaume2011}. It also presents one of the highest $\mathrm{\mathrm{T_c}}$ for an 
element\cite{PhysRevLett.91.167001,Struzhkin11082002,Schaeffer2015,PhysRevB.82.184509,shimizu:597,ashcroft:569} and it is expected to display a periodic undamped plasmon\cite{silkin:172102,PhysRevB.81.205105}. Additionally,
according to a recent experiment lithium shows quantum and isotope effects in its low temperature and pressure phase transformations\cite{Ackland2017}.

Experimental evidence\cite{Hanfland2000,Guillaume2011,matsuoka:186,1742-6596-121-5-052003,Schaeffernaturecomms,Ackland2017} shows that in the pressure and temperature ranges where the anomalous isotope effect was measured (15-25 GPa and below 30 K) lithium presents a fcc structure.
At around 40 GPa, it transforms to the rhombohedral hR1 phase, which is just a distortion of the fcc phase along the c axis if one switches to a hexagonal representation. The transformation to the cubic cI16 phase occurs shortly after, at around 43 GPa.

Theoretical calculations within the harmonic approximation in fcc lithium show a highly softened 
transverse acoustic mode in the $\Gamma K$ high-symmetry line\cite{PhysRevLett.96.047003,PhysRevLett.111.057006,PhysRevB.82.184509,doi:10.1143/JPSJ.74.3227,PhysRevB.74.172104}. Around $\mathrm{\mathbf{q}_{inst}}=2\pi/a (2/3,2/3,0)$,
where a is the lattice parameter, 
this anomalous mode presents a huge 
electron-phonon coupling, becoming a key factor to explain the high $\mathrm{\mathrm{T_c}}$ observed in lithium\cite{PhysRevLett.96.047003,PhysRevLett.111.057006,doi:10.1143/JPSJ.74.3227}. This softening is associated to a well defined Fermi surface 
nesting\cite{PhysRevLett.96.047003,PhysRevLett.111.057006,PhysRevB.82.184509,doi:10.1143/JPSJ.74.3227,PhysRevB.74.172104,rodriguez-prieto:125406}  and even yields imaginary phonon frequencies at pressures where fcc is known to be stable; the instability
emerges at pressures higher than 30 GPa in the local density approximation (LDA), 
and at even lower pressures if one uses the generalized gradient approximation (GGA). As seen in other systems, such as simple cubic Ca \cite{errea:112604}, PdH\cite{PhysRevLett.111.177002}, the record 
superconductor $\mathrm{H_3S}$\cite{Errea2016} and $\mathrm{NbSe_2}$\cite{PhysRevB.92.140303}, anharmonicity is expected to have a significant role stabilizing this structure and, 
due to phonon frequency renormalization, also determining its superconducting properties\cite{JJAPCP.6.011103}.
As it has been measured at lower pressures of the phase diagram of lithium\cite{Ackland2017}, zero-point vibrational energy could strongly impact the phase transitions of lithium in the 
15-25 GPa pressure range, specially considering the small enthalpy differences between the most competitive candidates according to previous calculations\cite{Hanfland2000,PhysRevB.78.014102,PhysRevB.79.054524}.
In fact, the anharmonic correction to the vibrational energy could be significant as well.   


\begin{figure}[t]
\includegraphics[width=\linewidth]{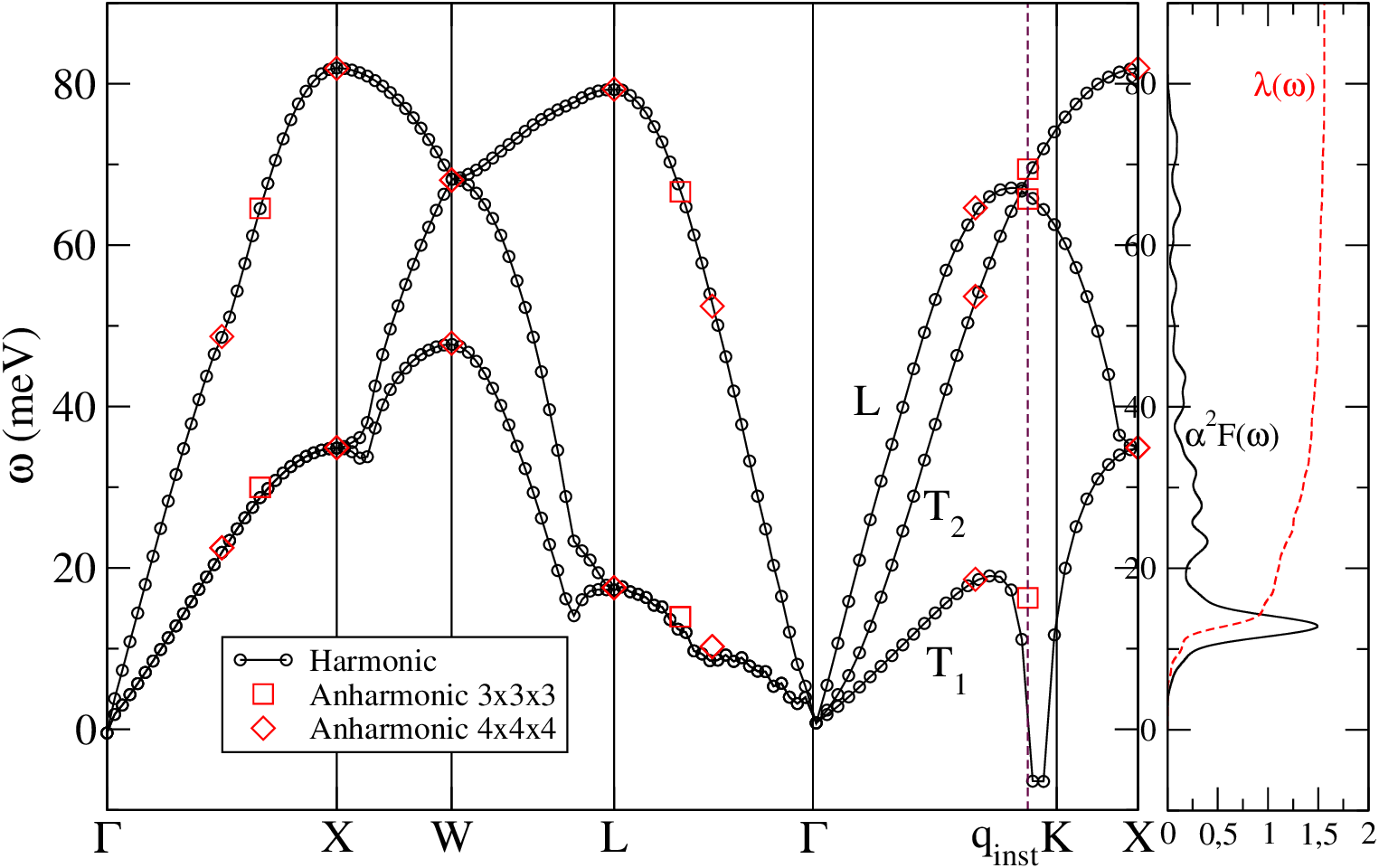}
\caption{Fcc $\mathrm{^7Li}$ phonon dispersion at 26 GPa. Anharmonic phonons within the SSCHA are calculated both for a $3\times3\times3$ and a $4\times4\times4$
grid of points. The Eliashberg function $\alpha^2F(\omega)$ and the integrated electron-phonon coupling $\lambda(\omega)$ is also shown for the anharmonic case.
\label{spectra_25_7}}
\end{figure}



The origin of the observed unconventional isotope effect in high pressure lithium remains unclear. Here we consider the following two hypothesis to explain this behavior. 
(i) Phonon frequencies scale with the atomic mass differently as expected within the harmonic approximation. Therefore, while in the harmonic approach the electron phonon coupling constant $\lambda$ is independent of the isotopic mass,
anharmonicity could make it differ from one isotope to the other, as it happens in palladium hydride \cite{PhysRevLett.111.177002}.
(ii) $\mathrm{^6Li}$ and $\mathrm{^7Li}$ isotopes adopt different crystal structures
due to the significant role of the vibrational energy in the phase diagram. Experimental evidence and previous theoretical calculations claim Li adopts the fcc phase from as low as 7 GPa to as high as 40 GPa in the
temperature regime where superconductivity has been measured\cite{Guillaume2011,matsuoka:186,1742-6596-121-5-052003,Schaeffernaturecomms}. However, there is a considerable lack of experimental data in the mentioned region of the phase diagram 
and all previous calculations have been done in the static approach.

In this work we present an exhaustive analysis of the superconducting properties of fcc and cI16 structures of lithium in the 15-45 GPa pressure range, with vibrational degrees of freedom 
 treated at the anharmonic level. We also analyze the possible existence of the hR1 phase in the pressure range of interest.

\begin{figure}[th]
\includegraphics[width=\linewidth]{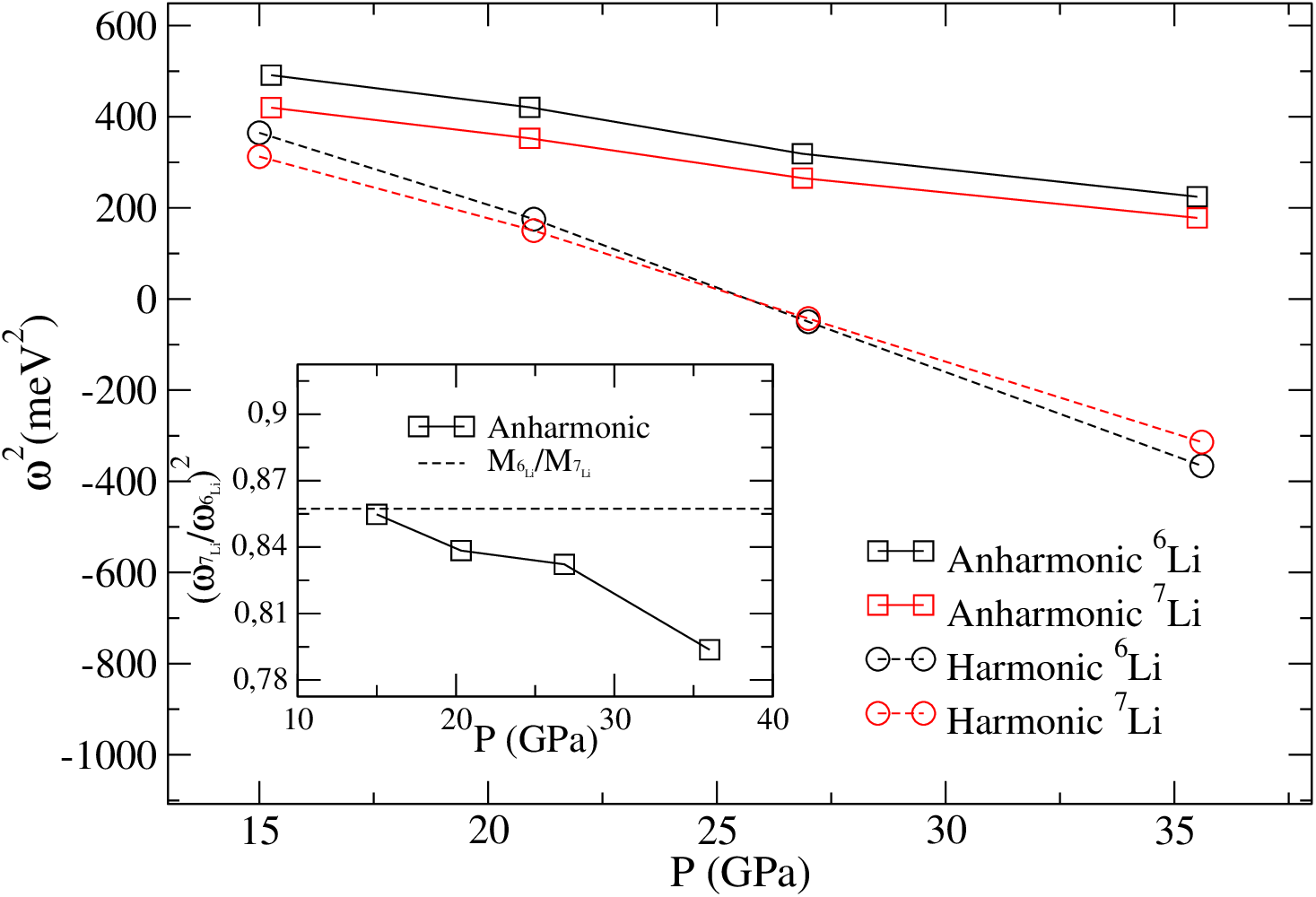}
\caption{Squared phonon frequencies of the anomalous transverse acoustic mode at $\mathrm{\mathbf{q}_{inst}}$ for $\mathrm{^6Li}$ and $\mathrm{^7Li}$ isotopes as a function of pressure.
The inset shows the ratio of the frequencies for both isotopes at the anharmonic level, $M_{\mathrm{^6Li}}/M_{\mathrm{^7Li}}$ being the harmonic value.\label{freqpress}}
\end{figure}

\section{Computational details}\label{computational}
Our density functional theory (DFT) calculations were done within the 
Perdew-Burke-Ernzerhof (PBE) parametrization of the GGA\cite{pbe}. Harmonic phonon frequencies and the electron-phonon deformation potential were calculated within density functional perturbation theory (DFPT)\cite{RevModPhys.73.515}
as implemented in {\sc Quantum ESPRESSO}\cite{0953-8984-21-39-395502}. The electron-proton
interaction was considered making use of an ultrasoft pseudopotential\cite{PhysRevB.41.7892} which includes 1s and 2s electrons. Anharmonic calculations, including the vibrational contribution to the enthalpy,
were performed using the stochastic self-consistent harmonic approximation (SSCHA)\cite{PhysRevB.89.064302}. Anharmonic force constant matrices of fcc lithium were obtained by calculating forces in $3\times3\times3$ supercells.
Therefore, anharmonic dynamical matrices were obtained in the respective commensurate \textbf{q}-point grids 
and interpolated to a finer $9\times9\times9$ mesh afterwards. These were combined with DFPT 
electron-phonon calculations obtained in the fine $9\times9\times9$ mesh to calculate the anharmonic Eliashberg 
function $\alpha^2F(\omega)$. The same procedure was used for the cI16 structure, being $2\times2\times2$ and $6\times6\times6$ the 
coarse and fine grids respectively. The vibrational contribution to the enthalpy of hR1, which is a distortion of the fcc phase, was calculated using a $2\times2\times2$ grid for obtaining anharmonic force constant matrices
and interpolating the differences with respect to the undistorted fcc structure. 
More details and the convergence parameters are given in the Supplementary Material.

\section{Results and Discussion}\label{results}

Fig. \ref{spectra_25_7} shows the DFPT harmonic phonon dispersion of fcc $\mathrm{^7Li}$ at 26 GPa and the anharmonic corrections calculated within the SSCHA. Anharmonic force constant matrices were obtained 
by calculating forces in  $3\times3\times3$ and $4\times4\times4$ supercells. Consequently, anharmonic dynamical matrices were obtained in the respective commensurate \textbf{q}-point grids.
We see that anharmonicity is primarily localized around the phonon softening at the transverse acoustic $T_1$ branch at $\mathrm{\mathbf{q}_{inst}}$, where the frequency is strongly shifted up by anharmonic effects.  
This well known phonon softening has been widely analyzed and explained in terms of Fermi surface nesting \cite{PhysRevLett.96.047003,PhysRevLett.111.057006,PhysRevB.82.184509,doi:10.1143/JPSJ.74.3227,PhysRevB.74.172104,rodriguez-prieto:125406}
 and, as shown in
Fig. \ref{freqpress}, it even yields imaginary frequencies at pressures higher than 25 GPa; a considerably lower pressure than the 30 GPa obtained within the LDA. 
In the same graph we also show the anharmonic frequency of the same mode, confirming fcc lithium is dynamically stabilized by anharmonicity above 25 GPa. 
However, as it is shown in the inset and even though this soft mode shows huge anharmonic effects, its frequency scales practically as in the harmonic case ( $\omega\propto \sqrt{1/M}$).
Despite the large anharmonicity, a similar harmonic scaling was previously calculated for high pressure simple cubic calcium\cite{errea:112604}. 

\begin{figure}[t]
\includegraphics[width=1\linewidth]{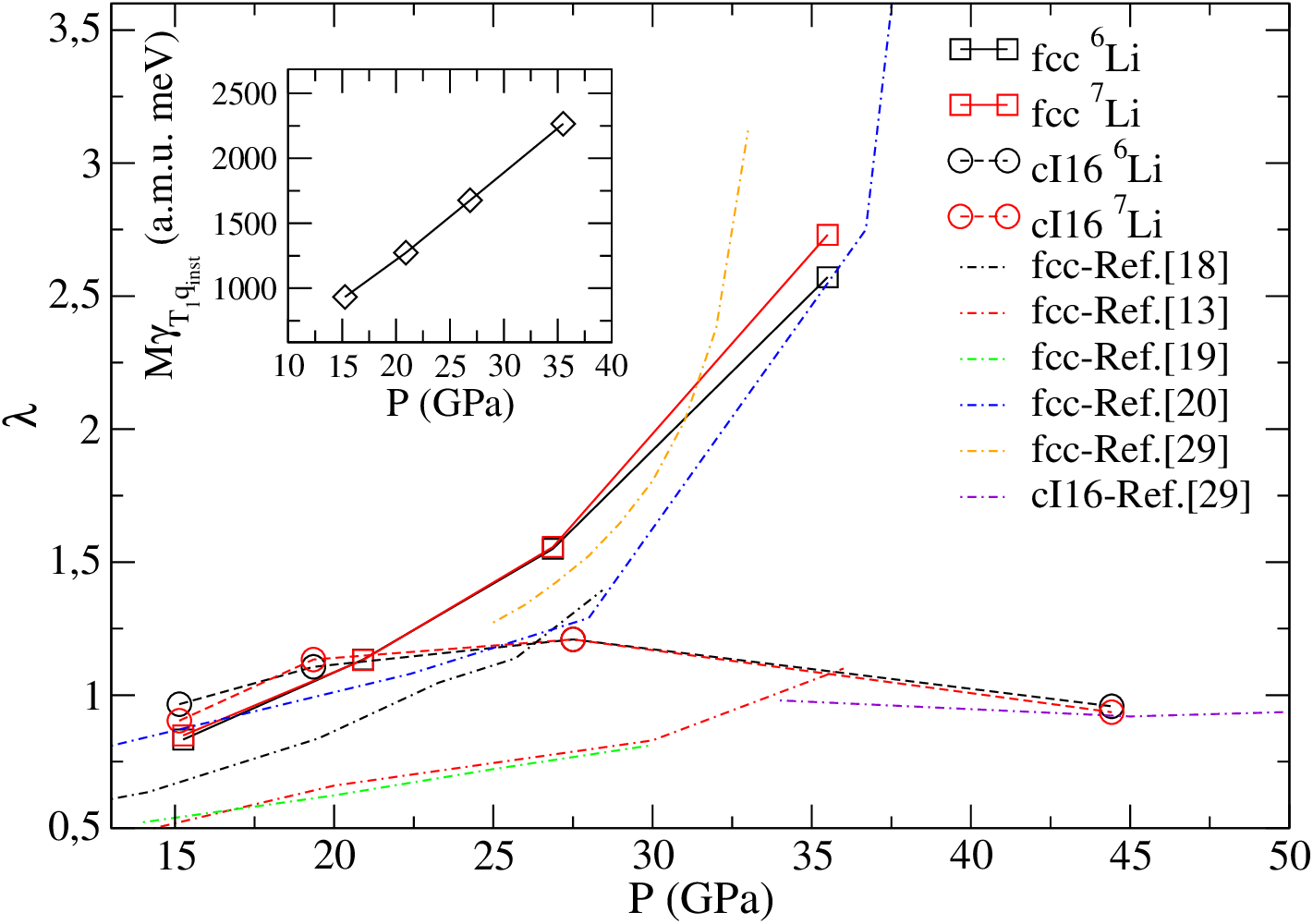}\caption{Total electron-phonon coupling constant $\lambda$ of fcc and cI16 lithium calculated for its two isotopes at different pressures.
The inset shows the phonon linewidth of the $T_1$ mode of fcc Li at $\mathrm{\mathbf{q}_{inst}}$ multiplied by the atomic mass, the product being independent of the phonon frequency and the isotopic mass.
the calculated $\lambda$ is compared to previous calculations\cite{PhysRevLett.96.047003,PhysRevLett.111.057006,PhysRevB.82.184509,doi:10.1143/JPSJ.74.3227,PhysRevB.79.054524}.\label{elphfig}}
\end{figure}

Our DFPT electron-phonon coupling calculations displayed in Fig. \ref{elphfig} show the total coupling constant $\lambda$ rises abruptly with increasing pressure in the fcc phase. Starting from 
an already high value of 0.85 at 15 GPa and reaching a value as high as 2.6 at 36 GPa, this dramatic growth is directly related to the also rapid increase of the electron-phonon linewidth $\gamma$ 
of the $T_1$ mode at $\mathrm{\mathbf{q}_{inst}}$, which doubles its value in the mentioned pressure range. The remarkable peak in the Eliashberg function $\alpha^2F(\omega)$ and the associated abrupt growth of the integrated electron-phonon coupling constant
$\lambda(\omega)$ around
the frequency of the anomaly is another indicator of how relevant this softening is in the superconducting properties of fcc lithium. However, while the phonon renormalization of the mentioned mode due to anharmonicity is huge, 
$\lambda$ is nearly identical for both isotopes at every pressure except at 35 GPa, where the difference is just 7\%, even if anharmonicity is already really strong. 
As mentioned above, this is due to the fact that the frequency of the anomalous mode scales harmonically.
Our $\lambda$ values are slightly larger than the ones by Maheswari \textit{et al.}\cite{doi:10.1143/JPSJ.74.3227} and Profeta \textit{et al.}\cite{PhysRevLett.96.047003} and quite larger than
the ones by Akashi \textit{et al.}\cite{PhysRevLett.111.057006} and Bazhirov \textit{et al.}\cite{PhysRevB.82.184509}. We attribute these disagreements to the large dependence of $\lambda$ with the \textbf{q}-point grid. 
While we used a $9\times9\times9$ sampling of the BZ for the electron-phonon and lattice dynamics calculations, where $\mathrm{\mathbf{q}_{inst}}$ is explicitly taken into account, the mentioned works use $8\times8\times8$ grids ($7\times7\times7$ in the case of Maheswari \textit{et al.}), where it is not. According to our convergence tests, 
those grids clearly underestimate $\lambda$ due to the absence of $\mathrm{\mathbf{q}_{inst}}$ in the grid (see Supplementary Material). Including this extremely anharmonic anomalous point is crucial for estimating the impact of anharmonicity
in the electron-phonon coupling and, as a consequence, the superconducting $\mathrm{T_c}$.

\begin{figure}[t]
\subfloat[(a)][Estimated $\mathrm{\mathrm{T_c}}$ of fcc and cI16 lithium for its two stable isotopes at different pressures (lines with symbols) and comparison with other theoretical estimations 
(dashed and dotted curves)~\cite{PhysRevLett.96.047003,PhysRevLett.111.057006,PhysRevB.82.184509,doi:10.1143/JPSJ.74.3227,PhysRevB.79.054524}.]{\includegraphics[width=1\linewidth]{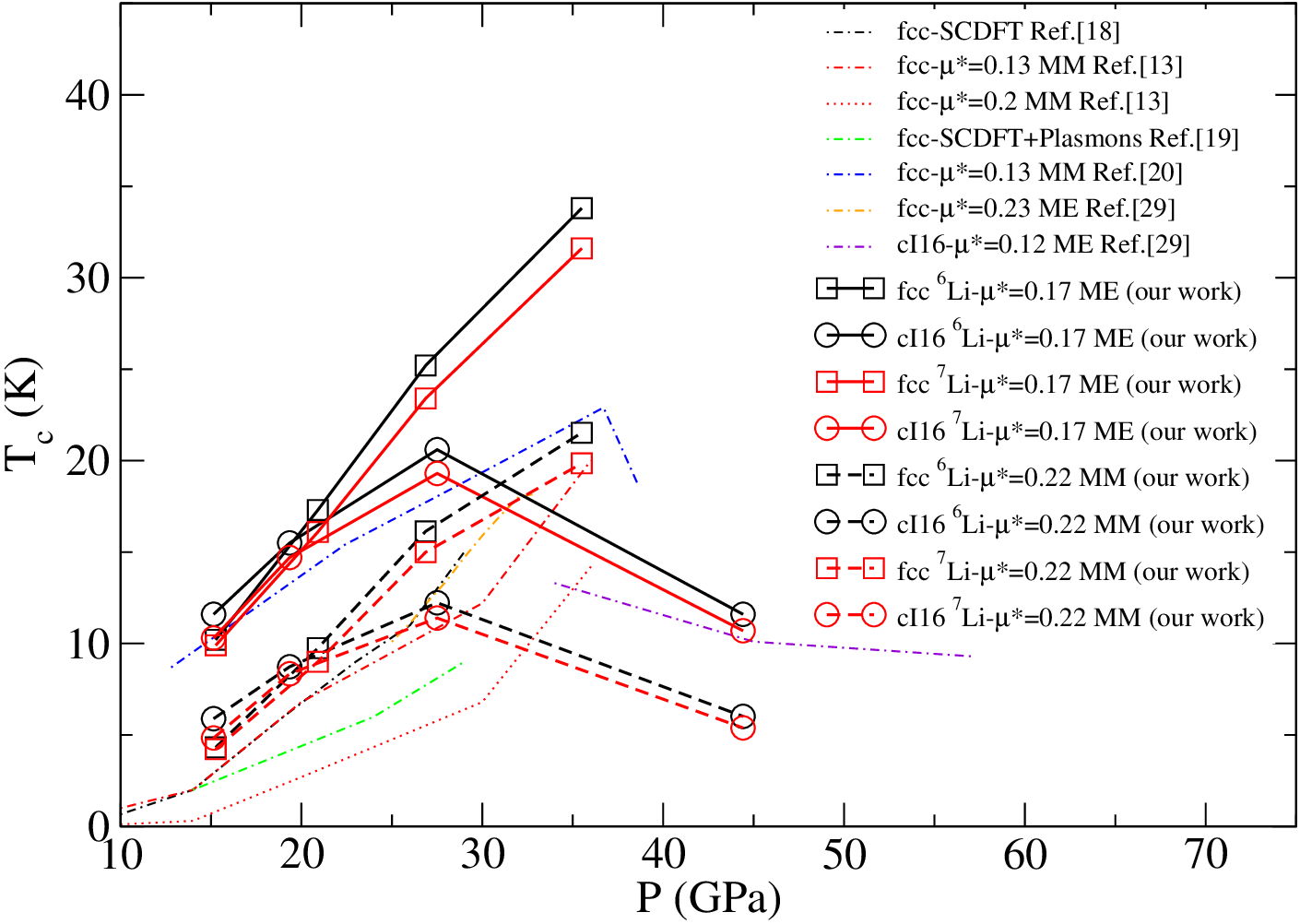}\label{theoreticalfig}}\\ 
\vspace{0.7cm} 
\subfloat[(b)][Estimated $\mathrm{\mathrm{T_c}}$ of fcc and cI16 lithium for its two stable isotopes at different pressures (lines with symbols) and comparison with experimental values (characters)
~\cite{Schaeffer2015,1742-6596-121-5-052003,PhysRevLett.91.167001,Struzhkin11082002,shimizu:597}.]
{\includegraphics[width=1\linewidth]{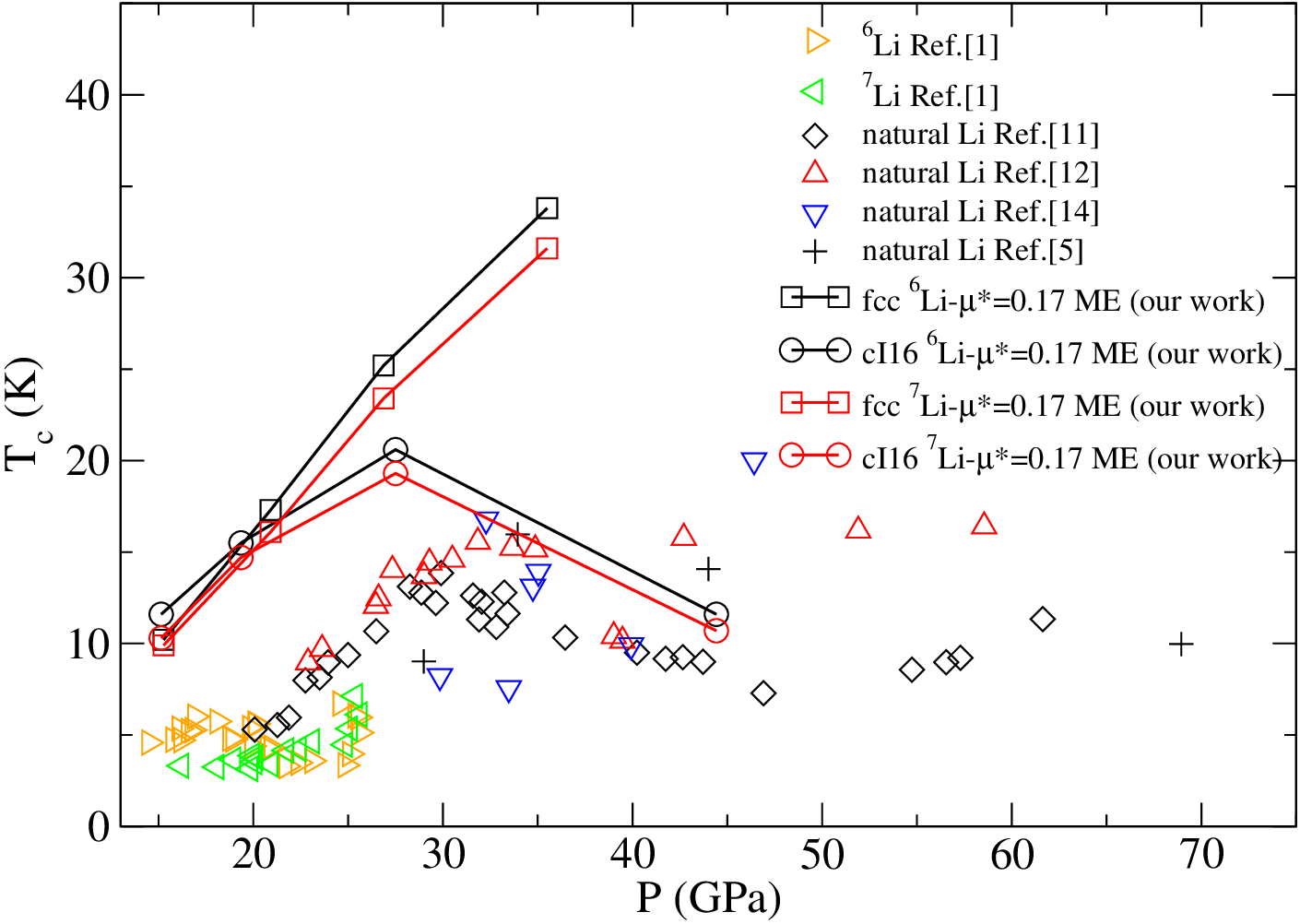}\label{expfig}}
\caption{$\mathrm{\mathrm{T_c}}$ estimations and comparison with (a) previous theoretical and (b) experimental results. \label{tcestimations}} 
\end{figure}

\begin{figure}[t]
\includegraphics[width=1\linewidth]{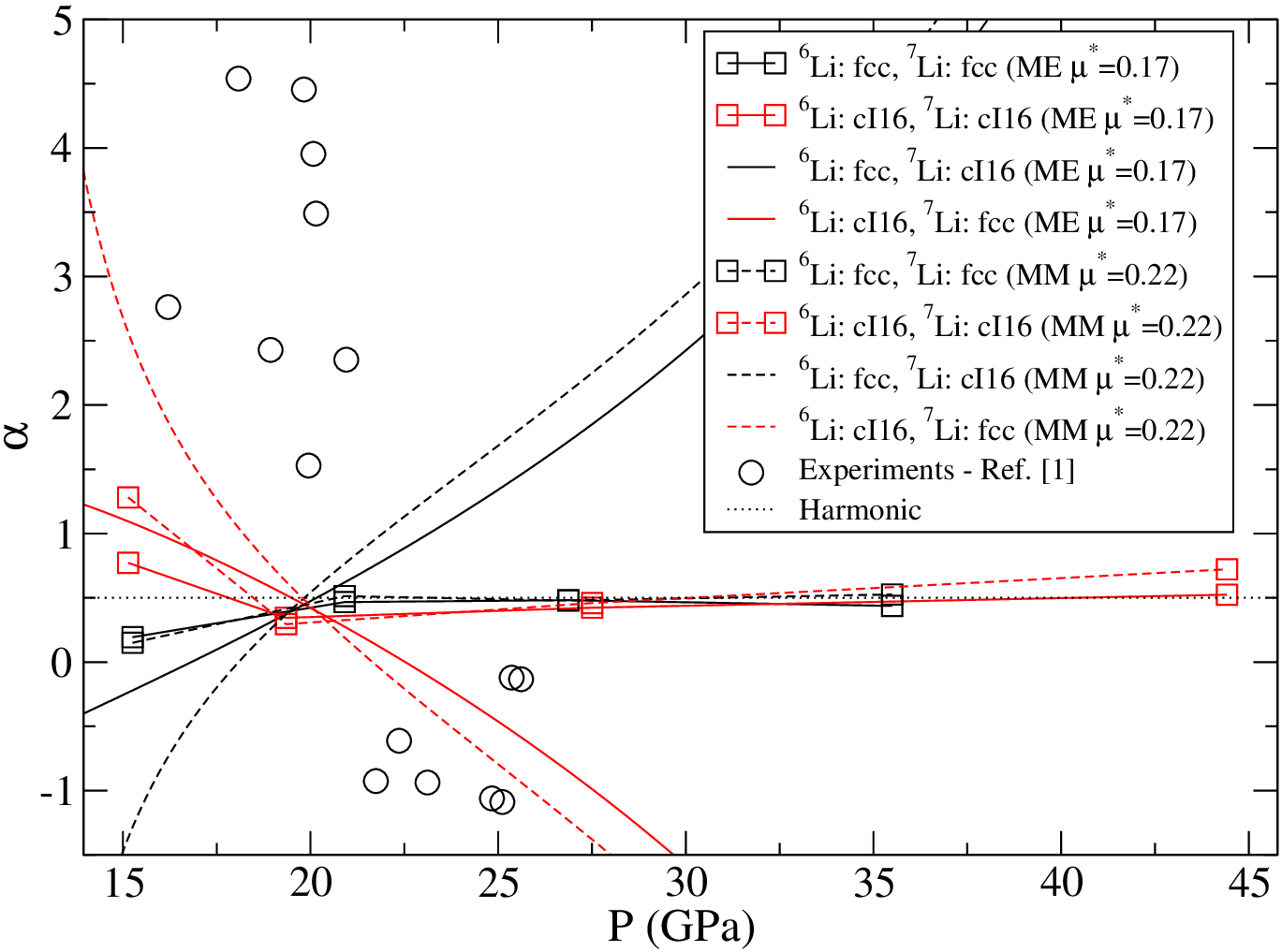}\caption{The isotope coefficient $\alpha$ against pressure. Lines with symbols show the coefficients obtained for the cases in which the two isotopes adopt the same crystal structure
(either cI16 or fcc). Curves without symbols show the coefficients for the cases in which the isotopes adopt different structures. \label{alpha}}
\end{figure}

Considering that for large electron-phonon coupling constants the McMillan equation underestimates the superconducting $\mathrm{\mathrm{T_c}}$\cite{PhysRevB.12.905},
we solved the isotropic Migdal-Eliashberg equations\cite{sov.phys.jetp.7.996,sov.phys.jetp.11.696}.
We estimated a $\mu^*$ value of 0.17 using the Morel-Anderson formula\cite{PhysRev.125.1263}:
\begin{equation}\label{mustar}
 \mu^*=\frac{\mu}{1+\ln{(\frac{\varepsilon_f}{\omega_D})}}.
\end{equation}
The average electron-electron Coulomb repulsion term $\mu$ was obtained from Thomas-Fermi screening theory, a free-electron Fermi energy $\varepsilon_f$ was chosen, and the Debye cutoff phonon frequency $\omega_D$ was 
taken as the highest frequency of the longitudinal acoustic modes \cite{PhysRevLett.54.2375}.  
Changes in phonon frequencies and electronic density for different pressures and isotopes only alter the fourth significant digit of $\mu^*$, so that differences in $\mu^*$ cannot explain the isotope effect anomalies and we assume the same value
for both isotopes. 
Fig. \ref{tcestimations} shows the superconducting critical temperature of fcc lithium for both isotopes at 15, 20, 26 and 36 GPa. We find $\mathrm{T_c}$ increases monotonically with pressure the same way $\lambda$ does, 
ranging from 11.2 K (10.7 K) at 15 GPa to 34.8 K (32.5 K) at 36 GPa for $\mathrm{^6Li}$ ($\mathrm{^7Li}$).
As in the case of $\lambda$, we do not see any anomalous scaling of the superconducting temperature with the isotopic mass; as it can be seen in Fig. \ref{alpha}, $\alpha$ is close to the conventional harmonic BCS value of 0.5 within the entire pressure range
except at 15 GPa where, even though it shows a lower value, it does not, in any case, explain the experimentally observed anomalous isotope effect.
Using McMillan's formula with $\mu^*=$0.22 $\mathrm{T_c}$ compares better with literature and experiments, even though values are still larger than in previous works due to the choice of the \textbf{q}-point grid as in the case of $\lambda$; 
in any case, $\alpha$ does not almost change, and the conclusion remains unaltered. The overestimation of $\mathrm{T_c}$ could also indicate that vertex corrections
in the electron-phonon coupling and anisotropic effects in the Migdal-Eliashberg
equations might be important. However, anisotropic effects should not be isotope dependent and, due to the harmonic scaling of phonon frequencies, we do not expect vertex corrections to yield any anomalous isotope effect either. 
Therefore, we discard hypothesis (i).

\begin{figure}[t]
\includegraphics[width=1\linewidth]{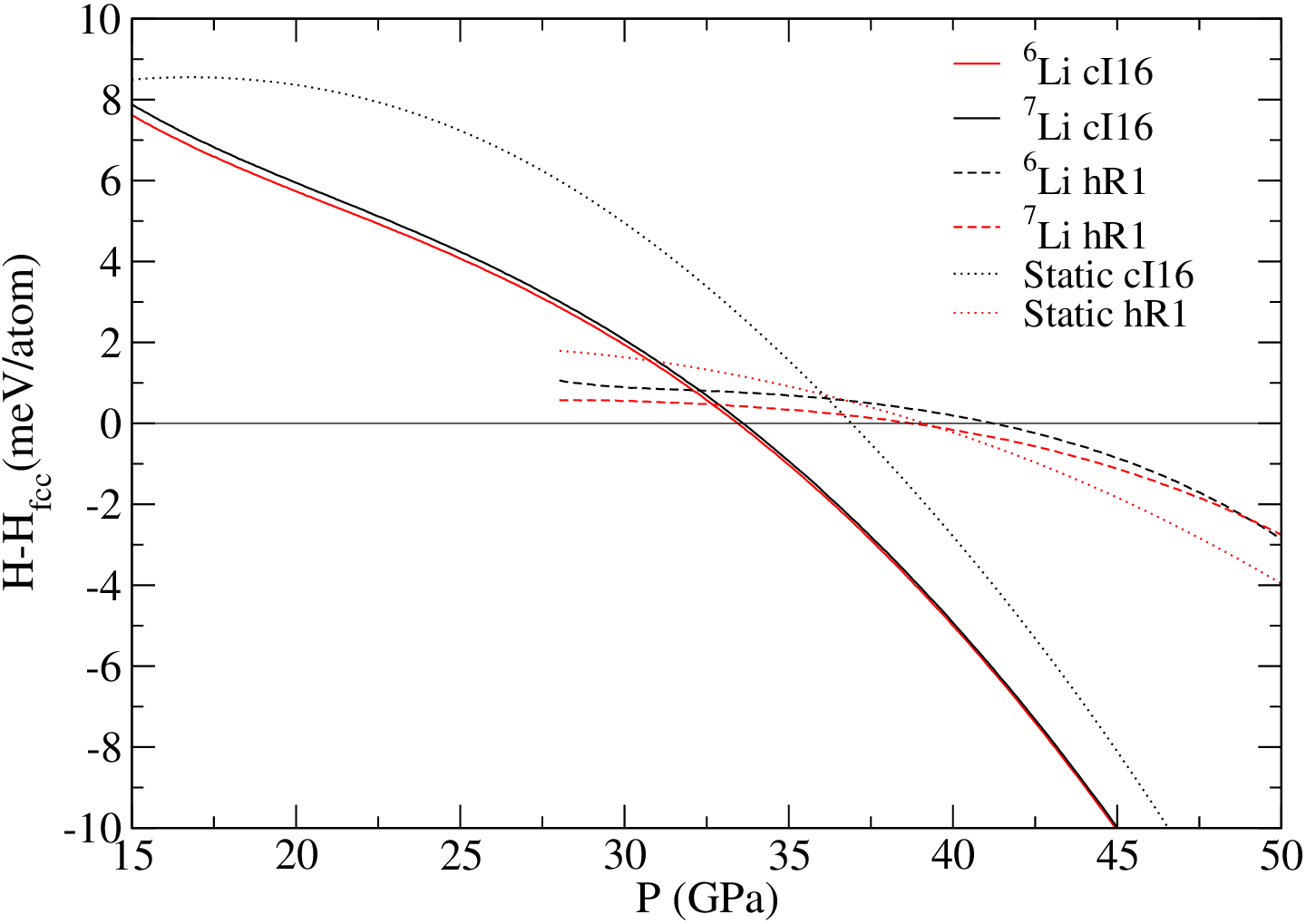}\caption{Relative enthalpies of cI16 and hR1 $\mathrm{^6Li}$ and $\mathrm{^7Li}$ isotopes with respect to their fcc counterparts. In solid and dashed lines ZPE has been included, while 
in dotted curves only electronic energy has been considered. The low pressure limit for the hR1 curves has been set at the pressure which corresponds, in each case, to the maximum volume at which the phase shows a local minimum 
in the total energy surface (see Supplementary Material).\label{phasetransition}}
\end{figure}

After discarding that the anomalous isotope effect comes from strong anharmonicity in the fcc phase, we analyzed the possibility of the two isotopes showing different structures at the same pressure in a thermodynamically stable way.
Fig. \ref{phasetransition} shows the enthalpies of the competing phases cI16 and hR1 relative to their respective fcc ones for the two isotopes. Our static calculations, \textit{i.e.} not including zero-point energy (ZPE),  compare well with literature 
(there are no previous works including ZPE)\cite{PhysRevB.78.014102} and just show the fcc to cI16 transition. No important changes are shown for both isotopes when anharmonic ZPE is included and, although in the pressure range where this phase transition happens the enthalpy difference with the hR1 is less than 1 meV per atom, that is, roughly the same as the error one assumes when converging total energy calculations within DFT, it remains metastable. Therefore, small changes in the calculation parameters or the choice of exchange and correlation potential might cause modifications in the transition pressures and phase sequence. Accordingly, when ZPE is included the fcc to cI16 transition pressure shifts from 37 GPa to 33 GPa for both isotopes, as the enthalpy difference is reduced by around 3 meV due to lattice
vibrations. Additionally, in the 21-25 GPa pressure range, where the inverse isotope effect was observed, the enthalpy difference between
cI16 and fcc structures is really small (around 4-6 meV/atom). In conclusion, our results do not support hypothesis (ii) as $\mathrm{^6Li}$ and $\mathrm{^7Li}$ isotopes are not expected to adopt different stable crystal structures. 
 
Due to the extremely small enthalpy differences metastable coexistence of phases can not be discarded as it happens at ambient pressure for its martensitic transition\cite{Ackland2017}. In order to see if $\mathrm{^6Li}$ and $\mathrm{^7Li}$ adopting different structures could lead to the observed anomalous isotope effect,
we have also made lattice dynamics and electron-phonon coupling calculations in the cI16 structure. We do not further consider hR1 as a candidate because, according to our calculations, the local minimum in the total energy surface associated to hR1 disappears for pressures lower than 28 GPa (see Supplementary Material). 
In Fig. \ref{elphfig} we show the total electron-phonon coupling $\lambda$ for cI16 Li at 15, 19, 27 and 44 GPa. $\lambda$ does not vary with pressure as much 
as it does in the fcc phase, it varies only between 0.9 and 1.2 in the 15-44 GPa pressure range. $\lambda$ is fairly similar for both isotopes, so that anharmonicity does not have almost any impact. Actually, at the lowest pressures, cI16 values differ more than the fcc ones from one isotope to the other. 
This is due to the fact that, while the overall phonon spectrum is very slightly modified by anharmonicity in the cI16 phase, anharmonic corrections occur mostly at the lowest frequencies, which are the ones that contribute most to 
the total electron-phonon coupling.    In fact, our $\lambda$ and $\mathrm{T_c}$ estimations, with $\mu^*$=0.17 obtained with the Morel-Anderson formula as in the fcc case, shown in Figs. \ref{elphfig} and \ref{tcestimations} yield values higher than in fcc below 20 GPa, being the opposite at higher pressures.
The isotope effect coefficient is close to the harmonic value at 27 and 44 GPa, with $\alpha=$0.42 and 0.57, respectively, while it deviates considerably from 0.5 at 15 and 19 GPa as it yields $\alpha=$0.77 and 0.34, respectively. 
All this agrees with the higher anharmonicity we found at lower pressures. 
Although our enthalpy calculations do not predict both isotopes can stabilize in different structures, we have also analyzed this metastability driven hypothetical scenario: $\mathrm{^6Li}$ stabilizing in the fcc phase and $\mathrm{^7Li}$ in the cI16,
and viceversa. 
As shown in Fig. \ref{alpha}, in the pressure range where the inverse isotope effect was experimentally observed (21-25 GPa), experimental values would only be qualitatively reproduced if $\mathrm{^6Li}$ adopted the cI16 structure while
$\mathrm{^7Li}$ were in the fcc phase.  This qualitative picture does not vary much if one uses the McMillan formula with $\mu^*=$0.22, but it could notably change if we used different $\mu^*$ values for the different phases.  

\section{Conclusions}

According to our calculations, even though anharmonicity is crucial to stabilize the fcc phase in lithium under pressure, its $\lambda$ remains almost the same for both isotopes and yields a conventional 
scaling of $\mathrm{\mathrm{T_c}}$ with isotopic mass and, therefore,  it does not explain  the experimentally observed anomalous isotope effect. On the other hand, including anharmonic ZPE in the enthalpy curve does 
not modify lithium phase diagram in the pressure range of interest, so that it is unexpected to have both isotopes in different structures. 
The anomalous isotope effect could only be qualitatively explained if $\mathrm{^7Li}$ adopted the fcc structure while $\mathrm{^6Li}$ adopted the cI16 one in a metastable way. All these, added to the large error bars and quite chaotic behavior of
$\mathrm{T_c}$ with pressure in Ref.~\onlinecite{Schaeffer2015}--with considerably different temperature values for the same pressure-- puts in question the experimental observation of an anomalous isotope effect in lithium at high pressure. This way, our work encourages further research to determine the
phase sequence and superconducting properties of the two stable isotopes of lithium. 

\section*{Acknowledgments}\label{acknowledgements}

The authors acknowledge financial support from the
Spanish Ministry of Economy and Competitiveness (FIS2016-76617-P) and the Department of Education, Universities and Research of the Basque 
Government and the University of the Basque Country (IT756-13).
M.B. is also thankful to the Department
of Education, Language Policy and Culture of the Basque
Government for a predoctoral fellowship (Grant No. PRE-2015-2-0269). 
Computer facilities were provided by PRACE and the 
Donostia International Physics Center (DIPC).

\bibliography{bibliografia}

\begin{thebibliography}{39}%
\makeatletter
\providecommand \@ifxundefined [1]{%
 \@ifx{#1\undefined}
}%
\providecommand \@ifnum [1]{%
 \ifnum #1\expandafter \@firstoftwo
 \else \expandafter \@secondoftwo
 \fi
}%
\providecommand \@ifx [1]{%
 \ifx #1\expandafter \@firstoftwo
 \else \expandafter \@secondoftwo
 \fi
}%
\providecommand \natexlab [1]{#1}%
\providecommand \enquote  [1]{``#1''}%
\providecommand \bibnamefont  [1]{#1}%
\providecommand \bibfnamefont [1]{#1}%
\providecommand \citenamefont [1]{#1}%
\providecommand \href@noop [0]{\@secondoftwo}%
\providecommand \href [0]{\begingroup \@sanitize@url \@href}%
\providecommand \@href[1]{\@@startlink{#1}\@@href}%
\providecommand \@@href[1]{\endgroup#1\@@endlink}%
\providecommand \@sanitize@url [0]{\catcode `\\12\catcode `\$12\catcode
  `\&12\catcode `\#12\catcode `\^12\catcode `\_12\catcode `\%12\relax}%
\providecommand \@@startlink[1]{}%
\providecommand \@@endlink[0]{}%
\providecommand \url  [0]{\begingroup\@sanitize@url \@url }%
\providecommand \@url [1]{\endgroup\@href {#1}{\urlprefix }}%
\providecommand \urlprefix  [0]{URL }%
\providecommand \Eprint [0]{\href }%
\providecommand \doibase [0]{http://dx.doi.org/}%
\providecommand \selectlanguage [0]{\@gobble}%
\providecommand \bibinfo  [0]{\@secondoftwo}%
\providecommand \bibfield  [0]{\@secondoftwo}%
\providecommand \translation [1]{[#1]}%
\providecommand \BibitemOpen [0]{}%
\providecommand \bibitemStop [0]{}%
\providecommand \bibitemNoStop [0]{.\EOS\space}%
\providecommand \EOS [0]{\spacefactor3000\relax}%
\providecommand \BibitemShut  [1]{\csname bibitem#1\endcsname}%
\let\auto@bib@innerbib\@empty
\bibitem [{\citenamefont {Schaeffer}\ \emph
  {et~al.}(2015{\natexlab{a}})\citenamefont {Schaeffer}, \citenamefont
  {Temple}, \citenamefont {Bishop},\ and\ \citenamefont
  {Deemyad}}]{Schaeffer2015}%
  \BibitemOpen
  \bibfield  {author} {\bibinfo {author} {\bibfnamefont {A.~M.}\ \bibnamefont
  {Schaeffer}}, \bibinfo {author} {\bibfnamefont {S.~R.}\ \bibnamefont
  {Temple}}, \bibinfo {author} {\bibfnamefont {J.~K.}\ \bibnamefont {Bishop}},
  \ and\ \bibinfo {author} {\bibfnamefont {S.}~\bibnamefont {Deemyad}},\ }\href
  {\doibase 10.1073/pnas.1412638112} {\bibfield  {journal} {\bibinfo  {journal}
  {Proceedings of the National Academy of Sciences}\ }\textbf {\bibinfo
  {volume} {112}},\ \bibinfo {pages} {60} (\bibinfo {year}
  {2015}{\natexlab{a}})}\BibitemShut {NoStop}%
\bibitem [{\citenamefont {Seitz}(1935)}]{Seitz1935}%
  \BibitemOpen
  \bibfield  {author} {\bibinfo {author} {\bibfnamefont {F.}~\bibnamefont
  {Seitz}},\ }\href {\doibase 10.1103/PhysRev.47.400} {\bibfield  {journal}
  {\bibinfo  {journal} {Physical Review}\ }\textbf {\bibinfo {volume} {47}},\
  \bibinfo {pages} {400} (\bibinfo {year} {1935})}\BibitemShut {NoStop}%
\bibitem [{\citenamefont {Guillaume}\ \emph {et~al.}(2011)\citenamefont
  {Guillaume}, \citenamefont {Gregoryanz}, \citenamefont {Degtyareva},
  \citenamefont {McMahon}, \citenamefont {Hanfland}, \citenamefont {Evans},
  \citenamefont {Guthrie}, \citenamefont {Sinogeikin},\ and\ \citenamefont
  {Mao}}]{Guillaume2011}%
  \BibitemOpen
  \bibfield  {author} {\bibinfo {author} {\bibfnamefont {C.~L.}\ \bibnamefont
  {Guillaume}}, \bibinfo {author} {\bibfnamefont {E.}~\bibnamefont
  {Gregoryanz}}, \bibinfo {author} {\bibfnamefont {O.}~\bibnamefont
  {Degtyareva}}, \bibinfo {author} {\bibfnamefont {M.~I.}\ \bibnamefont
  {McMahon}}, \bibinfo {author} {\bibfnamefont {M.}~\bibnamefont {Hanfland}},
  \bibinfo {author} {\bibfnamefont {S.}~\bibnamefont {Evans}}, \bibinfo
  {author} {\bibfnamefont {M.}~\bibnamefont {Guthrie}}, \bibinfo {author}
  {\bibfnamefont {S.~V.}\ \bibnamefont {Sinogeikin}}, \ and\ \bibinfo {author}
  {\bibfnamefont {H.-K.}\ \bibnamefont {Mao}},\ }\href {\doibase
  10.1038/nphys1864} {\bibfield  {journal} {\bibinfo  {journal} {Nature
  Physics}\ }\textbf {\bibinfo {volume} {7}},\ \bibinfo {pages} {211} (\bibinfo
  {year} {2011})}\BibitemShut {NoStop}%
\bibitem [{\citenamefont {Hanfland}\ \emph {et~al.}(2000)\citenamefont
  {Hanfland}, \citenamefont {Syassen}, \citenamefont {Christensen},\ and\
  \citenamefont {Novikov}}]{Hanfland2000}%
  \BibitemOpen
  \bibfield  {author} {\bibinfo {author} {\bibfnamefont {M.}~\bibnamefont
  {Hanfland}}, \bibinfo {author} {\bibfnamefont {K.}~\bibnamefont {Syassen}},
  \bibinfo {author} {\bibfnamefont {N.~E.}\ \bibnamefont {Christensen}}, \ and\
  \bibinfo {author} {\bibfnamefont {D.~L.}\ \bibnamefont {Novikov}},\ }\href
  {\doibase 10.1038/35041515} {\bibfield  {journal} {\bibinfo  {journal}
  {Nature}\ }\textbf {\bibinfo {volume} {408}},\ \bibinfo {pages} {174}
  (\bibinfo {year} {2000})}\BibitemShut {NoStop}%
\bibitem [{\citenamefont {Matsuoka}\ \emph {et~al.}(2008)\citenamefont
  {Matsuoka}, \citenamefont {Onoda}, \citenamefont {Kaneshige}, \citenamefont
  {Nakamoto}, \citenamefont {Shimizu}, \citenamefont {Kagayama},\ and\
  \citenamefont {Ohishi}}]{1742-6596-121-5-052003}%
  \BibitemOpen
  \bibfield  {author} {\bibinfo {author} {\bibfnamefont {T.}~\bibnamefont
  {Matsuoka}}, \bibinfo {author} {\bibfnamefont {S.}~\bibnamefont {Onoda}},
  \bibinfo {author} {\bibfnamefont {M.}~\bibnamefont {Kaneshige}}, \bibinfo
  {author} {\bibfnamefont {Y.}~\bibnamefont {Nakamoto}}, \bibinfo {author}
  {\bibfnamefont {K.}~\bibnamefont {Shimizu}}, \bibinfo {author} {\bibfnamefont
  {T.}~\bibnamefont {Kagayama}}, \ and\ \bibinfo {author} {\bibfnamefont
  {Y.}~\bibnamefont {Ohishi}},\ }\href
  {http://stacks.iop.org/1742-6596/121/i=5/a=052003} {\bibfield  {journal}
  {\bibinfo  {journal} {Journal of Physics: Conference Series}\ }\textbf
  {\bibinfo {volume} {121}},\ \bibinfo {pages} {052003} (\bibinfo {year}
  {2008})}\BibitemShut {NoStop}%
\bibitem [{\citenamefont {Schaeffer}\ \emph
  {et~al.}(2015{\natexlab{b}})\citenamefont {Schaeffer}, \citenamefont {Cai},
  \citenamefont {Olejnik}, \citenamefont {Molaison}, \citenamefont
  {Sinogeikin}, \citenamefont {dos Santos},\ and\ \citenamefont
  {Deemyad}}]{Schaeffernaturecomms}%
  \BibitemOpen
  \bibfield  {author} {\bibinfo {author} {\bibfnamefont {A.~M.}\ \bibnamefont
  {Schaeffer}}, \bibinfo {author} {\bibfnamefont {W.}~\bibnamefont {Cai}},
  \bibinfo {author} {\bibfnamefont {E.}~\bibnamefont {Olejnik}}, \bibinfo
  {author} {\bibfnamefont {J.~J.}\ \bibnamefont {Molaison}}, \bibinfo {author}
  {\bibfnamefont {S.}~\bibnamefont {Sinogeikin}}, \bibinfo {author}
  {\bibfnamefont {A.~M.}\ \bibnamefont {dos Santos}}, \ and\ \bibinfo {author}
  {\bibfnamefont {S.}~\bibnamefont {Deemyad}},\ }\href {\doibase
  10.1038/ncomms9030} {\bibfield  {journal} {\bibinfo  {journal} {Nature
  Communications}\ }\textbf {\bibinfo {volume} {6}},\ \bibinfo {pages} {8030}
  (\bibinfo {year} {2015}{\natexlab{b}})}\BibitemShut {NoStop}%
\bibitem [{\citenamefont {Ackland}\ \emph {et~al.}(2017)\citenamefont
  {Ackland}, \citenamefont {Dunuwille}, \citenamefont {Martinez-Canales},
  \citenamefont {Loa}, \citenamefont {Zhang}, \citenamefont {Sinogeikin},
  \citenamefont {Cai},\ and\ \citenamefont {Deemyad}}]{Ackland2017}%
  \BibitemOpen
  \bibfield  {author} {\bibinfo {author} {\bibfnamefont {G.~J.}\ \bibnamefont
  {Ackland}}, \bibinfo {author} {\bibfnamefont {M.}~\bibnamefont {Dunuwille}},
  \bibinfo {author} {\bibfnamefont {M.}~\bibnamefont {Martinez-Canales}},
  \bibinfo {author} {\bibfnamefont {I.}~\bibnamefont {Loa}}, \bibinfo {author}
  {\bibfnamefont {R.}~\bibnamefont {Zhang}}, \bibinfo {author} {\bibfnamefont
  {S.}~\bibnamefont {Sinogeikin}}, \bibinfo {author} {\bibfnamefont
  {W.}~\bibnamefont {Cai}}, \ and\ \bibinfo {author} {\bibfnamefont
  {S.}~\bibnamefont {Deemyad}},\ }\href {\doibase 10.1126/science.aal4886}
  {\bibfield  {journal} {\bibinfo  {journal} {Science}\ }\textbf {\bibinfo
  {volume} {356}},\ \bibinfo {pages} {1254} (\bibinfo {year}
  {2017})}\BibitemShut {NoStop}%
\bibitem [{\citenamefont {Rousseau}\ \emph {et~al.}(2011)\citenamefont
  {Rousseau}, \citenamefont {Xie}, \citenamefont {Ma},\ and\ \citenamefont
  {Bergara}}]{springerlink:10.1140/epjb/e2011-10972-9}%
  \BibitemOpen
  \bibfield  {author} {\bibinfo {author} {\bibfnamefont {B.}~\bibnamefont
  {Rousseau}}, \bibinfo {author} {\bibfnamefont {Y.}~\bibnamefont {Xie}},
  \bibinfo {author} {\bibfnamefont {Y.}~\bibnamefont {Ma}}, \ and\ \bibinfo
  {author} {\bibfnamefont {A.}~\bibnamefont {Bergara}},\ }\href
  {http://dx.doi.org/10.1140/epjb/e2011-10972-9} {\bibfield  {journal}
  {\bibinfo  {journal} {Eur. Phys. J. B}\ }\textbf {\bibinfo {volume} {81}},\
  \bibinfo {pages} {1} (\bibinfo {year} {2011})}\BibitemShut {NoStop}%
\bibitem [{\citenamefont {Matsuoka}\ and\ \citenamefont
  {Shimizu}(2009)}]{matsuoka:186}%
  \BibitemOpen
  \bibfield  {author} {\bibinfo {author} {\bibfnamefont {T.}~\bibnamefont
  {Matsuoka}}\ and\ \bibinfo {author} {\bibfnamefont {K.}~\bibnamefont
  {Shimizu}},\ }\href {\doibase 10.1038/nature07827} {\bibfield  {journal}
  {\bibinfo  {journal} {Nature}\ }\textbf {\bibinfo {volume} {458}},\ \bibinfo
  {pages} {186} (\bibinfo {year} {2009})}\BibitemShut {NoStop}%
\bibitem [{\citenamefont {Hern\'andez}\ \emph {et~al.}(2010)\citenamefont
  {Hern\'andez}, \citenamefont {Rodriguez-Prieto}, \citenamefont {Bergara},\
  and\ \citenamefont {Alf\`e}}]{PhysRevLett.104.185701}%
  \BibitemOpen
  \bibfield  {author} {\bibinfo {author} {\bibfnamefont {E.~R.}\ \bibnamefont
  {Hern\'andez}}, \bibinfo {author} {\bibfnamefont {A.}~\bibnamefont
  {Rodriguez-Prieto}}, \bibinfo {author} {\bibfnamefont {A.}~\bibnamefont
  {Bergara}}, \ and\ \bibinfo {author} {\bibfnamefont {D.}~\bibnamefont
  {Alf\`e}},\ }\href {\doibase 10.1103/PhysRevLett.104.185701} {\bibfield
  {journal} {\bibinfo  {journal} {Phys. Rev. Lett.}\ }\textbf {\bibinfo
  {volume} {104}},\ \bibinfo {pages} {185701} (\bibinfo {year}
  {2010})}\BibitemShut {NoStop}%
\bibitem [{\citenamefont {Deemyad}\ and\ \citenamefont
  {Schilling}(2003)}]{PhysRevLett.91.167001}%
  \BibitemOpen
  \bibfield  {author} {\bibinfo {author} {\bibfnamefont {S.}~\bibnamefont
  {Deemyad}}\ and\ \bibinfo {author} {\bibfnamefont {J.~S.}\ \bibnamefont
  {Schilling}},\ }\href {\doibase 10.1103/PhysRevLett.91.167001} {\bibfield
  {journal} {\bibinfo  {journal} {Phys. Rev. Lett.}\ }\textbf {\bibinfo
  {volume} {91}},\ \bibinfo {pages} {167001} (\bibinfo {year}
  {2003})}\BibitemShut {NoStop}%
\bibitem [{\citenamefont {Struzhkin}\ \emph {et~al.}(2002)\citenamefont
  {Struzhkin}, \citenamefont {Eremets}, \citenamefont {Gan}, \citenamefont
  {Mao},\ and\ \citenamefont {Hemley}}]{Struzhkin11082002}%
  \BibitemOpen
  \bibfield  {author} {\bibinfo {author} {\bibfnamefont {V.~V.}\ \bibnamefont
  {Struzhkin}}, \bibinfo {author} {\bibfnamefont {M.~I.}\ \bibnamefont
  {Eremets}}, \bibinfo {author} {\bibfnamefont {W.}~\bibnamefont {Gan}},
  \bibinfo {author} {\bibfnamefont {H.-k.}\ \bibnamefont {Mao}}, \ and\
  \bibinfo {author} {\bibfnamefont {R.~J.}\ \bibnamefont {Hemley}},\ }\href
  {\doibase 10.1126/science.1078535} {\bibfield  {journal} {\bibinfo  {journal}
  {Science}\ }\textbf {\bibinfo {volume} {298}},\ \bibinfo {pages} {1213}
  (\bibinfo {year} {2002})}\BibitemShut {NoStop}%
\bibitem [{\citenamefont {Bazhirov}\ \emph {et~al.}(2010)\citenamefont
  {Bazhirov}, \citenamefont {Noffsinger},\ and\ \citenamefont
  {Cohen}}]{PhysRevB.82.184509}%
  \BibitemOpen
  \bibfield  {author} {\bibinfo {author} {\bibfnamefont {T.}~\bibnamefont
  {Bazhirov}}, \bibinfo {author} {\bibfnamefont {J.}~\bibnamefont
  {Noffsinger}}, \ and\ \bibinfo {author} {\bibfnamefont {M.~L.}\ \bibnamefont
  {Cohen}},\ }\href {\doibase 10.1103/PhysRevB.82.184509} {\bibfield  {journal}
  {\bibinfo  {journal} {Phys. Rev. B}\ }\textbf {\bibinfo {volume} {82}},\
  \bibinfo {pages} {184509} (\bibinfo {year} {2010})}\BibitemShut {NoStop}%
\bibitem [{\citenamefont {Shimizu}\ \emph {et~al.}(2002)\citenamefont
  {Shimizu}, \citenamefont {Ishikawa}, \citenamefont {Takao}, \citenamefont
  {Yagi},\ and\ \citenamefont {Amaya}}]{shimizu:597}%
  \BibitemOpen
  \bibfield  {author} {\bibinfo {author} {\bibfnamefont {K.}~\bibnamefont
  {Shimizu}}, \bibinfo {author} {\bibfnamefont {H.}~\bibnamefont {Ishikawa}},
  \bibinfo {author} {\bibfnamefont {D.}~\bibnamefont {Takao}}, \bibinfo
  {author} {\bibfnamefont {T.}~\bibnamefont {Yagi}}, \ and\ \bibinfo {author}
  {\bibfnamefont {K.}~\bibnamefont {Amaya}},\ }\href {\doibase
  10.1038/nature01098} {\bibfield  {journal} {\bibinfo  {journal} {Nature}\
  }\textbf {\bibinfo {volume} {419}},\ \bibinfo {pages} {597} (\bibinfo {year}
  {2002})}\BibitemShut {NoStop}%
\bibitem [{\citenamefont {Ashcroft}(2002)}]{ashcroft:569}%
  \BibitemOpen
  \bibfield  {author} {\bibinfo {author} {\bibfnamefont {N.~W.}\ \bibnamefont
  {Ashcroft}},\ }\href {\doibase 10.1038/419569a} {\bibfield  {journal}
  {\bibinfo  {journal} {Nature}\ }\textbf {\bibinfo {volume} {419}},\ \bibinfo
  {pages} {569} (\bibinfo {year} {2002})}\BibitemShut {NoStop}%
\bibitem [{\citenamefont {Silkin}\ \emph {et~al.}(2007)\citenamefont {Silkin},
  \citenamefont {Rodriguez-Prieto}, \citenamefont {Bergara}, \citenamefont
  {Chulkov},\ and\ \citenamefont {Echenique}}]{silkin:172102}%
  \BibitemOpen
  \bibfield  {author} {\bibinfo {author} {\bibfnamefont {V.~M.}\ \bibnamefont
  {Silkin}}, \bibinfo {author} {\bibfnamefont {A.}~\bibnamefont
  {Rodriguez-Prieto}}, \bibinfo {author} {\bibfnamefont {A.}~\bibnamefont
  {Bergara}}, \bibinfo {author} {\bibfnamefont {E.~V.}\ \bibnamefont
  {Chulkov}}, \ and\ \bibinfo {author} {\bibfnamefont {P.~M.}\ \bibnamefont
  {Echenique}},\ }\href {\doibase 10.1103/PhysRevB.75.172102} {\bibfield
  {journal} {\bibinfo  {journal} {Phys. Rev. B}\ }\textbf {\bibinfo {volume}
  {75}},\ \bibinfo {eid} {172102} (\bibinfo {year} {2007})}\BibitemShut
  {NoStop}%
\bibitem [{\citenamefont {Errea}\ \emph {et~al.}(2010)\citenamefont {Errea},
  \citenamefont {Rodriguez-Prieto}, \citenamefont {Rousseau}, \citenamefont
  {Silkin},\ and\ \citenamefont {Bergara}}]{PhysRevB.81.205105}%
  \BibitemOpen
  \bibfield  {author} {\bibinfo {author} {\bibfnamefont {I.}~\bibnamefont
  {Errea}}, \bibinfo {author} {\bibfnamefont {A.}~\bibnamefont
  {Rodriguez-Prieto}}, \bibinfo {author} {\bibfnamefont {B.}~\bibnamefont
  {Rousseau}}, \bibinfo {author} {\bibfnamefont {V.~M.}\ \bibnamefont
  {Silkin}}, \ and\ \bibinfo {author} {\bibfnamefont {A.}~\bibnamefont
  {Bergara}},\ }\href {\doibase 10.1103/PhysRevB.81.205105} {\bibfield
  {journal} {\bibinfo  {journal} {Phys. Rev. B}\ }\textbf {\bibinfo {volume}
  {81}},\ \bibinfo {pages} {205105} (\bibinfo {year} {2010})}\BibitemShut
  {NoStop}%
\bibitem [{\citenamefont {Profeta}\ \emph {et~al.}(2006)\citenamefont
  {Profeta}, \citenamefont {Franchini}, \citenamefont {Lathiotakis},
  \citenamefont {Floris}, \citenamefont {Sanna}, \citenamefont {Marques},
  \citenamefont {L\"uders}, \citenamefont {Massidda}, \citenamefont {Gross},\
  and\ \citenamefont {Continenza}}]{PhysRevLett.96.047003}%
  \BibitemOpen
  \bibfield  {author} {\bibinfo {author} {\bibfnamefont {G.}~\bibnamefont
  {Profeta}}, \bibinfo {author} {\bibfnamefont {C.}~\bibnamefont {Franchini}},
  \bibinfo {author} {\bibfnamefont {N.~N.}\ \bibnamefont {Lathiotakis}},
  \bibinfo {author} {\bibfnamefont {A.}~\bibnamefont {Floris}}, \bibinfo
  {author} {\bibfnamefont {A.}~\bibnamefont {Sanna}}, \bibinfo {author}
  {\bibfnamefont {M.~A.~L.}\ \bibnamefont {Marques}}, \bibinfo {author}
  {\bibfnamefont {M.}~\bibnamefont {L\"uders}}, \bibinfo {author}
  {\bibfnamefont {S.}~\bibnamefont {Massidda}}, \bibinfo {author}
  {\bibfnamefont {E.~K.~U.}\ \bibnamefont {Gross}}, \ and\ \bibinfo {author}
  {\bibfnamefont {A.}~\bibnamefont {Continenza}},\ }\href {\doibase
  10.1103/PhysRevLett.96.047003} {\bibfield  {journal} {\bibinfo  {journal}
  {Phys. Rev. Lett.}\ }\textbf {\bibinfo {volume} {96}},\ \bibinfo {pages}
  {047003} (\bibinfo {year} {2006})}\BibitemShut {NoStop}%
\bibitem [{\citenamefont {Akashi}\ and\ \citenamefont
  {Arita}(2013)}]{PhysRevLett.111.057006}%
  \BibitemOpen
  \bibfield  {author} {\bibinfo {author} {\bibfnamefont {R.}~\bibnamefont
  {Akashi}}\ and\ \bibinfo {author} {\bibfnamefont {R.}~\bibnamefont {Arita}},\
  }\href {\doibase 10.1103/PhysRevLett.111.057006} {\bibfield  {journal}
  {\bibinfo  {journal} {Phys. Rev. Lett.}\ }\textbf {\bibinfo {volume} {111}},\
  \bibinfo {pages} {057006} (\bibinfo {year} {2013})}\BibitemShut {NoStop}%
\bibitem [{\citenamefont {Maheswari}\ \emph {et~al.}(2005)\citenamefont
  {Maheswari}, \citenamefont {Nagara}, \citenamefont {Kusakabe},\ and\
  \citenamefont {Suzuki}}]{doi:10.1143/JPSJ.74.3227}%
  \BibitemOpen
  \bibfield  {author} {\bibinfo {author} {\bibfnamefont {S.~U.}\ \bibnamefont
  {Maheswari}}, \bibinfo {author} {\bibfnamefont {H.}~\bibnamefont {Nagara}},
  \bibinfo {author} {\bibfnamefont {K.}~\bibnamefont {Kusakabe}}, \ and\
  \bibinfo {author} {\bibfnamefont {N.}~\bibnamefont {Suzuki}},\ }\href
  {\doibase 10.1143/JPSJ.74.3227} {\bibfield  {journal} {\bibinfo  {journal}
  {Journal of the Physical Society of Japan}\ }\textbf {\bibinfo {volume}
  {74}},\ \bibinfo {pages} {3227} (\bibinfo {year} {2005})}\BibitemShut
  {NoStop}%
\bibitem [{\citenamefont {Rodriguez-Prieto}\ \emph {et~al.}(2006)\citenamefont
  {Rodriguez-Prieto}, \citenamefont {Bergara}, \citenamefont {Silkin},\ and\
  \citenamefont {Echenique}}]{PhysRevB.74.172104}%
  \BibitemOpen
  \bibfield  {author} {\bibinfo {author} {\bibfnamefont {A.}~\bibnamefont
  {Rodriguez-Prieto}}, \bibinfo {author} {\bibfnamefont {A.}~\bibnamefont
  {Bergara}}, \bibinfo {author} {\bibfnamefont {V.~M.}\ \bibnamefont {Silkin}},
  \ and\ \bibinfo {author} {\bibfnamefont {P.~M.}\ \bibnamefont {Echenique}},\
  }\href {\doibase 10.1103/PhysRevB.74.172104} {\bibfield  {journal} {\bibinfo
  {journal} {Phys. Rev. B}\ }\textbf {\bibinfo {volume} {74}},\ \bibinfo
  {pages} {172104} (\bibinfo {year} {2006})}\BibitemShut {NoStop}%
\bibitem [{\citenamefont {Rodriguez-Prieto}\ and\ \citenamefont
  {Bergara}(2005)}]{rodriguez-prieto:125406}%
  \BibitemOpen
  \bibfield  {author} {\bibinfo {author} {\bibfnamefont {A.}~\bibnamefont
  {Rodriguez-Prieto}}\ and\ \bibinfo {author} {\bibfnamefont {A.}~\bibnamefont
  {Bergara}},\ }\href {\doibase 10.1103/PhysRevB.72.125406} {\bibfield
  {journal} {\bibinfo  {journal} {Phys. Rev. B}\ }\textbf {\bibinfo {volume}
  {72}},\ \bibinfo {pages} {125406} (\bibinfo {year} {2005})}\BibitemShut
  {NoStop}%
\bibitem [{\citenamefont {Errea}\ \emph {et~al.}(2012)\citenamefont {Errea},
  \citenamefont {Rousseau},\ and\ \citenamefont {Bergara}}]{errea:112604}%
  \BibitemOpen
  \bibfield  {author} {\bibinfo {author} {\bibfnamefont {I.}~\bibnamefont
  {Errea}}, \bibinfo {author} {\bibfnamefont {B.}~\bibnamefont {Rousseau}}, \
  and\ \bibinfo {author} {\bibfnamefont {A.}~\bibnamefont {Bergara}},\ }\href
  {\doibase 10.1063/1.4726161} {\bibfield  {journal} {\bibinfo  {journal}
  {Journal of Applied Physics}\ }\textbf {\bibinfo {volume} {111}},\ \bibinfo
  {eid} {112604} (\bibinfo {year} {2012})}\BibitemShut {NoStop}%
\bibitem [{\citenamefont {Errea}\ \emph {et~al.}(2013)\citenamefont {Errea},
  \citenamefont {Calandra},\ and\ \citenamefont
  {Mauri}}]{PhysRevLett.111.177002}%
  \BibitemOpen
  \bibfield  {author} {\bibinfo {author} {\bibfnamefont {I.}~\bibnamefont
  {Errea}}, \bibinfo {author} {\bibfnamefont {M.}~\bibnamefont {Calandra}}, \
  and\ \bibinfo {author} {\bibfnamefont {F.}~\bibnamefont {Mauri}},\ }\href
  {\doibase 10.1103/PhysRevLett.111.177002} {\bibfield  {journal} {\bibinfo
  {journal} {Phys. Rev. Lett.}\ }\textbf {\bibinfo {volume} {111}},\ \bibinfo
  {pages} {177002} (\bibinfo {year} {2013})}\BibitemShut {NoStop}%
\bibitem [{\citenamefont {Errea}\ \emph {et~al.}(2016)\citenamefont {Errea},
  \citenamefont {Calandra}, \citenamefont {Pickard}, \citenamefont {Nelson},
  \citenamefont {Needs}, \citenamefont {Li}, \citenamefont {Liu}, \citenamefont
  {Zhang}, \citenamefont {Ma},\ and\ \citenamefont {Mauri}}]{Errea2016}%
  \BibitemOpen
  \bibfield  {author} {\bibinfo {author} {\bibfnamefont {I.}~\bibnamefont
  {Errea}}, \bibinfo {author} {\bibfnamefont {M.}~\bibnamefont {Calandra}},
  \bibinfo {author} {\bibfnamefont {C.~J.}\ \bibnamefont {Pickard}}, \bibinfo
  {author} {\bibfnamefont {J.~R.}\ \bibnamefont {Nelson}}, \bibinfo {author}
  {\bibfnamefont {R.~J.}\ \bibnamefont {Needs}}, \bibinfo {author}
  {\bibfnamefont {Y.}~\bibnamefont {Li}}, \bibinfo {author} {\bibfnamefont
  {H.}~\bibnamefont {Liu}}, \bibinfo {author} {\bibfnamefont {Y.}~\bibnamefont
  {Zhang}}, \bibinfo {author} {\bibfnamefont {Y.}~\bibnamefont {Ma}}, \ and\
  \bibinfo {author} {\bibfnamefont {F.}~\bibnamefont {Mauri}},\ }\href
  {\doibase 10.1038/nature17175} {\bibfield  {journal} {\bibinfo  {journal}
  {Nature}\ }\textbf {\bibinfo {volume} {532}},\ \bibinfo {pages} {81}
  (\bibinfo {year} {2016})}\BibitemShut {NoStop}%
\bibitem [{\citenamefont {Leroux}\ \emph {et~al.}(2015)\citenamefont {Leroux},
  \citenamefont {Errea}, \citenamefont {Le~Tacon}, \citenamefont {Souliou},
  \citenamefont {Garbarino}, \citenamefont {Cario}, \citenamefont {Bosak},
  \citenamefont {Mauri}, \citenamefont {Calandra},\ and\ \citenamefont
  {Rodi\`ere}}]{PhysRevB.92.140303}%
  \BibitemOpen
  \bibfield  {author} {\bibinfo {author} {\bibfnamefont {M.}~\bibnamefont
  {Leroux}}, \bibinfo {author} {\bibfnamefont {I.}~\bibnamefont {Errea}},
  \bibinfo {author} {\bibfnamefont {M.}~\bibnamefont {Le~Tacon}}, \bibinfo
  {author} {\bibfnamefont {S.-M.}\ \bibnamefont {Souliou}}, \bibinfo {author}
  {\bibfnamefont {G.}~\bibnamefont {Garbarino}}, \bibinfo {author}
  {\bibfnamefont {L.}~\bibnamefont {Cario}}, \bibinfo {author} {\bibfnamefont
  {A.}~\bibnamefont {Bosak}}, \bibinfo {author} {\bibfnamefont
  {F.}~\bibnamefont {Mauri}}, \bibinfo {author} {\bibfnamefont
  {M.}~\bibnamefont {Calandra}}, \ and\ \bibinfo {author} {\bibfnamefont
  {P.}~\bibnamefont {Rodi\`ere}},\ }\href {\doibase 10.1103/PhysRevB.92.140303}
  {\bibfield  {journal} {\bibinfo  {journal} {Phys. Rev. B}\ }\textbf {\bibinfo
  {volume} {92}},\ \bibinfo {pages} {140303} (\bibinfo {year}
  {2015})}\BibitemShut {NoStop}%
\bibitem [{\citenamefont {Borinaga}\ \emph {et~al.}(2017)\citenamefont
  {Borinaga}, \citenamefont {Aseginolaza}, \citenamefont {Errea}, \citenamefont
  {Bergara}, \citenamefont {Aseginolaza}, \citenamefont {Errea},\ and\
  \citenamefont {Bergara}}]{JJAPCP.6.011103}%
  \BibitemOpen
  \bibfield  {author} {\bibinfo {author} {\bibfnamefont {M.}~\bibnamefont
  {Borinaga}}, \bibinfo {author} {\bibfnamefont {U.}~\bibnamefont
  {Aseginolaza}}, \bibinfo {author} {\bibfnamefont {I.}~\bibnamefont {Errea}},
  \bibinfo {author} {\bibfnamefont {A.}~\bibnamefont {Bergara}}, \bibinfo
  {author} {\bibfnamefont {U.}~\bibnamefont {Aseginolaza}}, \bibinfo {author}
  {\bibfnamefont {I.}~\bibnamefont {Errea}}, \ and\ \bibinfo {author}
  {\bibfnamefont {A.}~\bibnamefont {Bergara}},\ }\href {\doibase
  10.7567/JJAPCP.6.011103} {\bibfield  {journal} {\bibinfo  {journal} {JJAP
  Conference Proceedings}\ }\textbf {\bibinfo {volume} {011103}},\ \bibinfo
  {pages} {6} (\bibinfo {year} {2017})}\BibitemShut {NoStop}%
\bibitem [{\citenamefont {Ma}\ \emph {et~al.}(2008)\citenamefont {Ma},
  \citenamefont {Oganov},\ and\ \citenamefont {Xie}}]{PhysRevB.78.014102}%
  \BibitemOpen
  \bibfield  {author} {\bibinfo {author} {\bibfnamefont {Y.}~\bibnamefont
  {Ma}}, \bibinfo {author} {\bibfnamefont {A.~R.}\ \bibnamefont {Oganov}}, \
  and\ \bibinfo {author} {\bibfnamefont {Y.}~\bibnamefont {Xie}},\ }\href
  {\doibase 10.1103/PhysRevB.78.014102} {\bibfield  {journal} {\bibinfo
  {journal} {Phys. Rev. B}\ }\textbf {\bibinfo {volume} {78}},\ \bibinfo
  {pages} {014102} (\bibinfo {year} {2008})}\BibitemShut {NoStop}%
\bibitem [{\citenamefont {Yao}\ \emph {et~al.}(2009)\citenamefont {Yao},
  \citenamefont {Tse}, \citenamefont {Tanaka}, \citenamefont {Marsiglio},\ and\
  \citenamefont {Ma}}]{PhysRevB.79.054524}%
  \BibitemOpen
  \bibfield  {author} {\bibinfo {author} {\bibfnamefont {Y.}~\bibnamefont
  {Yao}}, \bibinfo {author} {\bibfnamefont {J.~S.}\ \bibnamefont {Tse}},
  \bibinfo {author} {\bibfnamefont {K.}~\bibnamefont {Tanaka}}, \bibinfo
  {author} {\bibfnamefont {F.}~\bibnamefont {Marsiglio}}, \ and\ \bibinfo
  {author} {\bibfnamefont {Y.}~\bibnamefont {Ma}},\ }\href {\doibase
  10.1103/PhysRevB.79.054524} {\bibfield  {journal} {\bibinfo  {journal} {Phys.
  Rev. B}\ }\textbf {\bibinfo {volume} {79}},\ \bibinfo {pages} {054524}
  (\bibinfo {year} {2009})}\BibitemShut {NoStop}%
\bibitem [{\citenamefont {Perdew}\ \emph {et~al.}(1996)\citenamefont {Perdew},
  \citenamefont {Burke},\ and\ \citenamefont {Ernzerhof}}]{pbe}%
  \BibitemOpen
  \bibfield  {author} {\bibinfo {author} {\bibfnamefont {J.~P.}\ \bibnamefont
  {Perdew}}, \bibinfo {author} {\bibfnamefont {K.}~\bibnamefont {Burke}}, \
  and\ \bibinfo {author} {\bibfnamefont {M.}~\bibnamefont {Ernzerhof}},\ }\href
  {\doibase 10.1103/PhysRevLett.77.3865} {\bibfield  {journal} {\bibinfo
  {journal} {Phys. Rev. Lett.}\ }\textbf {\bibinfo {volume} {77}},\ \bibinfo
  {pages} {3865} (\bibinfo {year} {1996})}\BibitemShut {NoStop}%
\bibitem [{\citenamefont {Baroni}\ \emph {et~al.}(2001)\citenamefont {Baroni},
  \citenamefont {de~Gironcoli}, \citenamefont {Dal~Corso},\ and\ \citenamefont
  {Giannozzi}}]{RevModPhys.73.515}%
  \BibitemOpen
  \bibfield  {author} {\bibinfo {author} {\bibfnamefont {S.}~\bibnamefont
  {Baroni}}, \bibinfo {author} {\bibfnamefont {S.}~\bibnamefont
  {de~Gironcoli}}, \bibinfo {author} {\bibfnamefont {A.}~\bibnamefont
  {Dal~Corso}}, \ and\ \bibinfo {author} {\bibfnamefont {P.}~\bibnamefont
  {Giannozzi}},\ }\href {\doibase 10.1103/RevModPhys.73.515} {\bibfield
  {journal} {\bibinfo  {journal} {Rev. Mod. Phys.}\ }\textbf {\bibinfo {volume}
  {73}},\ \bibinfo {pages} {515} (\bibinfo {year} {2001})}\BibitemShut
  {NoStop}%
\bibitem [{\citenamefont {Giannozzi~\emph{et
  al.}}(2009)}]{0953-8984-21-39-395502}%
  \BibitemOpen
  \bibfield  {author} {\bibinfo {author} {\bibfnamefont {P.}~\bibnamefont
  {Giannozzi~\emph{et al.}}},\ }\href {\doibase 10.1088/0953-8984/21/39/395502}
  {\bibfield  {journal} {\bibinfo  {journal} {J. Phys. Condens. Matter}\
  }\textbf {\bibinfo {volume} {21}},\ \bibinfo {pages} {395502} (\bibinfo
  {year} {2009})}\BibitemShut {NoStop}%
\bibitem [{\citenamefont {Vanderbilt}(1990)}]{PhysRevB.41.7892}%
  \BibitemOpen
  \bibfield  {author} {\bibinfo {author} {\bibfnamefont {D.}~\bibnamefont
  {Vanderbilt}},\ }\href {\doibase 10.1103/PhysRevB.41.7892} {\bibfield
  {journal} {\bibinfo  {journal} {Phys. Rev. B}\ }\textbf {\bibinfo {volume}
  {41}},\ \bibinfo {pages} {7892} (\bibinfo {year} {1990})}\BibitemShut
  {NoStop}%
\bibitem [{\citenamefont {Errea}\ \emph {et~al.}(2014)\citenamefont {Errea},
  \citenamefont {Calandra},\ and\ \citenamefont {Mauri}}]{PhysRevB.89.064302}%
  \BibitemOpen
  \bibfield  {author} {\bibinfo {author} {\bibfnamefont {I.}~\bibnamefont
  {Errea}}, \bibinfo {author} {\bibfnamefont {M.}~\bibnamefont {Calandra}}, \
  and\ \bibinfo {author} {\bibfnamefont {F.}~\bibnamefont {Mauri}},\ }\href
  {\doibase 10.1103/PhysRevB.89.064302} {\bibfield  {journal} {\bibinfo
  {journal} {Phys. Rev. B}\ }\textbf {\bibinfo {volume} {89}},\ \bibinfo
  {pages} {064302} (\bibinfo {year} {2014})}\BibitemShut {NoStop}%
\bibitem [{\citenamefont {Allen}\ and\ \citenamefont
  {Dynes}(1975)}]{PhysRevB.12.905}%
  \BibitemOpen
  \bibfield  {author} {\bibinfo {author} {\bibfnamefont {P.~B.}\ \bibnamefont
  {Allen}}\ and\ \bibinfo {author} {\bibfnamefont {R.~C.}\ \bibnamefont
  {Dynes}},\ }\href {\doibase 10.1103/PhysRevB.12.905} {\bibfield  {journal}
  {\bibinfo  {journal} {Phys. Rev. B}\ }\textbf {\bibinfo {volume} {12}},\
  \bibinfo {pages} {905} (\bibinfo {year} {1975})}\BibitemShut {NoStop}%
\bibitem [{\citenamefont {Migdal}(1958)}]{sov.phys.jetp.7.996}%
  \BibitemOpen
  \bibfield  {author} {\bibinfo {author} {\bibfnamefont {A.~B.}\ \bibnamefont
  {Migdal}},\ }\href {http://www.jetp.ac.ru/cgi-bin/e/index/e/7/6/p996?a=list}
  {\bibfield  {journal} {\bibinfo  {journal} {Sov. Phys. JETP}\ }\textbf
  {\bibinfo {volume} {7}},\ \bibinfo {pages} {996} (\bibinfo {year}
  {1958})}\BibitemShut {NoStop}%
\bibitem [{\citenamefont {Eliashberg}(1960)}]{sov.phys.jetp.11.696}%
  \BibitemOpen
  \bibfield  {author} {\bibinfo {author} {\bibfnamefont {G.~M.}\ \bibnamefont
  {Eliashberg}},\ }\href
  {http://www.jetp.ac.ru/cgi-bin/e/index/e/11/3/p696?a=list} {\bibfield
  {journal} {\bibinfo  {journal} {Sov. Phys. JETP}\ }\textbf {\bibinfo {volume}
  {11}},\ \bibinfo {pages} {696} (\bibinfo {year} {1960})}\BibitemShut
  {NoStop}%
\bibitem [{\citenamefont {Morel}\ and\ \citenamefont
  {Anderson}(1962)}]{PhysRev.125.1263}%
  \BibitemOpen
  \bibfield  {author} {\bibinfo {author} {\bibfnamefont {P.}~\bibnamefont
  {Morel}}\ and\ \bibinfo {author} {\bibfnamefont {P.~W.}\ \bibnamefont
  {Anderson}},\ }\href {\doibase 10.1103/PhysRev.125.1263} {\bibfield
  {journal} {\bibinfo  {journal} {Phys. Rev.}\ }\textbf {\bibinfo {volume}
  {125}},\ \bibinfo {pages} {1263} (\bibinfo {year} {1962})}\BibitemShut
  {NoStop}%
\bibitem [{\citenamefont {Chang}\ \emph {et~al.}(1985)\citenamefont {Chang},
  \citenamefont {Dacorogna}, \citenamefont {Cohen}, \citenamefont {Mignot},
  \citenamefont {Chouteau},\ and\ \citenamefont
  {Martinez}}]{PhysRevLett.54.2375}%
  \BibitemOpen
  \bibfield  {author} {\bibinfo {author} {\bibfnamefont {K.~J.}\ \bibnamefont
  {Chang}}, \bibinfo {author} {\bibfnamefont {M.~M.}\ \bibnamefont
  {Dacorogna}}, \bibinfo {author} {\bibfnamefont {M.~L.}\ \bibnamefont
  {Cohen}}, \bibinfo {author} {\bibfnamefont {J.~M.}\ \bibnamefont {Mignot}},
  \bibinfo {author} {\bibfnamefont {G.}~\bibnamefont {Chouteau}}, \ and\
  \bibinfo {author} {\bibfnamefont {G.}~\bibnamefont {Martinez}},\ }\href
  {\doibase 10.1103/PhysRevLett.54.2375} {\bibfield  {journal} {\bibinfo
  {journal} {Phys. Rev. Lett.}\ }\textbf {\bibinfo {volume} {54}},\ \bibinfo
  {pages} {2375} (\bibinfo {year} {1985})}\BibitemShut {NoStop}%
\end{thebibliography}%


\begin{thebibliography}{2}%
\makeatletter
\providecommand \@ifxundefined [1]{%
 \@ifx{#1\undefined}
}%
\providecommand \@ifnum [1]{%
 \ifnum #1\expandafter \@firstoftwo
 \else \expandafter \@secondoftwo
 \fi
}%
\providecommand \@ifx [1]{%
 \ifx #1\expandafter \@firstoftwo
 \else \expandafter \@secondoftwo
 \fi
}%
\providecommand \natexlab [1]{#1}%
\providecommand \enquote  [1]{``#1''}%
\providecommand \bibnamefont  [1]{#1}%
\providecommand \bibfnamefont [1]{#1}%
\providecommand \citenamefont [1]{#1}%
\providecommand \href@noop [0]{\@secondoftwo}%
\providecommand \href [0]{\begingroup \@sanitize@url \@href}%
\providecommand \@href[1]{\@@startlink{#1}\@@href}%
\providecommand \@@href[1]{\endgroup#1\@@endlink}%
\providecommand \@sanitize@url [0]{\catcode `\\12\catcode `\$12\catcode
  `\&12\catcode `\#12\catcode `\^12\catcode `\_12\catcode `\%12\relax}%
\providecommand \@@startlink[1]{}%
\providecommand \@@endlink[0]{}%
\providecommand \url  [0]{\begingroup\@sanitize@url \@url }%
\providecommand \@url [1]{\endgroup\@href {#1}{\urlprefix }}%
\providecommand \urlprefix  [0]{URL }%
\providecommand \Eprint [0]{\href }%
\providecommand \doibase [0]{http://dx.doi.org/}%
\providecommand \selectlanguage [0]{\@gobble}%
\providecommand \bibinfo  [0]{\@secondoftwo}%
\providecommand \bibfield  [0]{\@secondoftwo}%
\providecommand \translation [1]{[#1]}%
\providecommand \BibitemOpen [0]{}%
\providecommand \bibitemStop [0]{}%
\providecommand \bibitemNoStop [0]{.\EOS\space}%
\providecommand \EOS [0]{\spacefactor3000\relax}%
\providecommand \BibitemShut  [1]{\csname bibitem#1\endcsname}%
\let\auto@bib@innerbib\@empty
\bibitem [{\citenamefont {Vanderbilt}(1990)}]{PhysRevB.41.7892}%
  \BibitemOpen
  \bibfield  {author} {\bibinfo {author} {\bibfnamefont {D.}~\bibnamefont
  {Vanderbilt}},\ }\href {\doibase 10.1103/PhysRevB.41.7892} {\bibfield
  {journal} {\bibinfo  {journal} {Phys. Rev. B}\ }\textbf {\bibinfo {volume}
  {41}},\ \bibinfo {pages} {7892} (\bibinfo {year} {1990})}\BibitemShut
  {NoStop}%
\bibitem [{\citenamefont {Errea}\ \emph {et~al.}(2014)\citenamefont {Errea},
  \citenamefont {Calandra},\ and\ \citenamefont {Mauri}}]{PhysRevB.89.064302}%
  \BibitemOpen
  \bibfield  {author} {\bibinfo {author} {\bibfnamefont {I.}~\bibnamefont
  {Errea}}, \bibinfo {author} {\bibfnamefont {M.}~\bibnamefont {Calandra}}, \
  and\ \bibinfo {author} {\bibfnamefont {F.}~\bibnamefont {Mauri}},\ }\href
  {\doibase 10.1103/PhysRevB.89.064302} {\bibfield  {journal} {\bibinfo
  {journal} {Phys. Rev. B}\ }\textbf {\bibinfo {volume} {89}},\ \bibinfo
  {pages} {064302} (\bibinfo {year} {2014})}\BibitemShut {NoStop}%
\end{thebibliography}%

\end{document}


\selectlanguage{USenglish}



\title{ \textbf{Supplementary Material} \\~ \\ Anharmonicity and the isotope effect in superconducting lithium at high pressures: a first-principles approach}

\author[1,2]{Miguel Borinaga}
\author[1,2]{Unai Aseginolaza}
\author[2,3]{Ion Errea}
\author[4]{Matteo Calandra}
\author[5]{Francesco Mauri}
\author[1,2,6]{Aitor Bergara}

\affil[1]{Centro de F\'isica de Materiales CFM, CSIC-UPV/EHU, Paseo Manuel de
             Lardizabal 5, 20018 Donostia/San Sebasti\'an, Basque Country, Spain}
\affil[2]{Donostia International Physics Center
             (DIPC), Manuel Lardizabal pasealekua 4, 20018 Donostia/San
             Sebasti\'an, Basque Country, Spain}
\affil[3]{Fisika Aplikatua 1 Saila, Bilboko Ingeniaritza Eskola,
             University of the Basque Country (UPV/EHU), Rafael Moreno ``Pitxitxi'' Pasealekua 3, 48013 Bilbao,
             Basque Country, Spain}
\affil[4]{IMPMC, UMR CNRS 7590, Sorbonne
Universit\'es - UPMC Univ. Paris 06, MNHN, IRD, 4 Place Jussieu,
F-75005 Paris, France}
\affil[5]{Dipartimento di Fisica, Universit\`a di Roma La Sapienza, Piazzale Aldo Moro 5, I-00185 Roma, Italy}
\affil[6]{Departamento de F\'isica de la Materia Condensada,  University of the Basque Country (UPV/EHU), 48080 Bilbao, 
             Basque Country, Spain}

\renewcommand\Authands{ and }
\renewcommand*{\Affilfont}{\small\itshape}

\date{\vspace{-5ex}}
\maketitle

\section{Phonon spectra and electron-phonon coupling}

\begin{figure}[t]
 \includegraphics[width=0.8\linewidth]{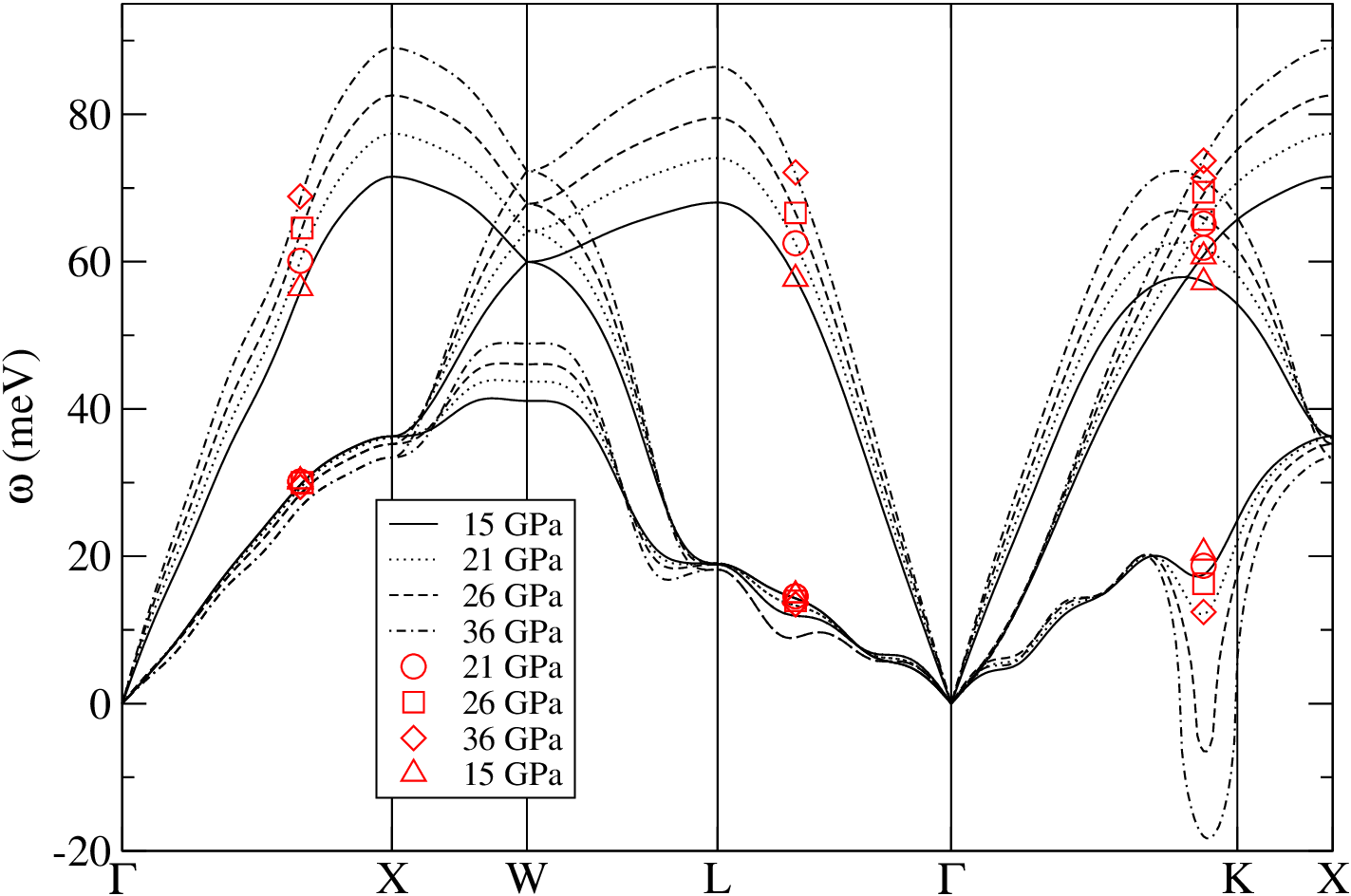}
 \caption{Harmonic (black curves) and anharmonic (red symbols) phonon spectra of fcc Li at different pressures. Harmonic spectra are obtained by Fourier interpolating the $9\times9\times9$ \textbf{q}-grid data to the desired path. 
Anharmonic data corresponds to SSCHA calculations in a $3\times3\times3$ \textbf{q}-point grid.\label{fccphononspectra}}
\end{figure}

\subsection{fcc structure}

Harmonic dynamical matrices of fcc Li have been obtained in a $9\times9\times9$ \textbf{q}-grid for every analyzed pressure and isotope. A proper convergence of phonon frequencies required a $30\times30\times30$ \textbf{k}-point grid and a Methfessel-Paxton
smearing width of 0.01 Ry for electronic integrations in the first BZ. An energy cutoff of 65 Ry was necessary for expanding the wave-functions in the plane-wave basis. The electron-proton
interaction was considered making use of an ultrasoft pseudopotential\cite{PhysRevB.41.7892}, in which $1s^2$ core electrons where also included (the same pseudopotential has been used for the whole work). In Fig. \ref{fccphononspectra} we show the harmonic phonon spectra obtained at 
15,21,26 and 36 GPa for $\mathrm{^7Li}$ after Fourier interpolating the dynamical matrices from the $9\times9\times9$ grid to the desired path. 

%

Anharmonic dynamical matrices where obtained in a $3\times3\times3$ \textbf{q}-grid commensurate to the supercell size for our SSCHA calculations, in which we calculate forces acting on atoms. The difference between
the harmonic and anharmonic dynamical matrices was interpolated to the finer $9\times9\times9$ grid, so that anharmonic dynamical matrices were obtained in $9\times9\times9$ grid after adding the harmonic dynamical matrix to the interpolation. 


\begin{figure}[th]
\subfloat[(a)][15 GPa]{\includegraphics[width=0.48\linewidth]{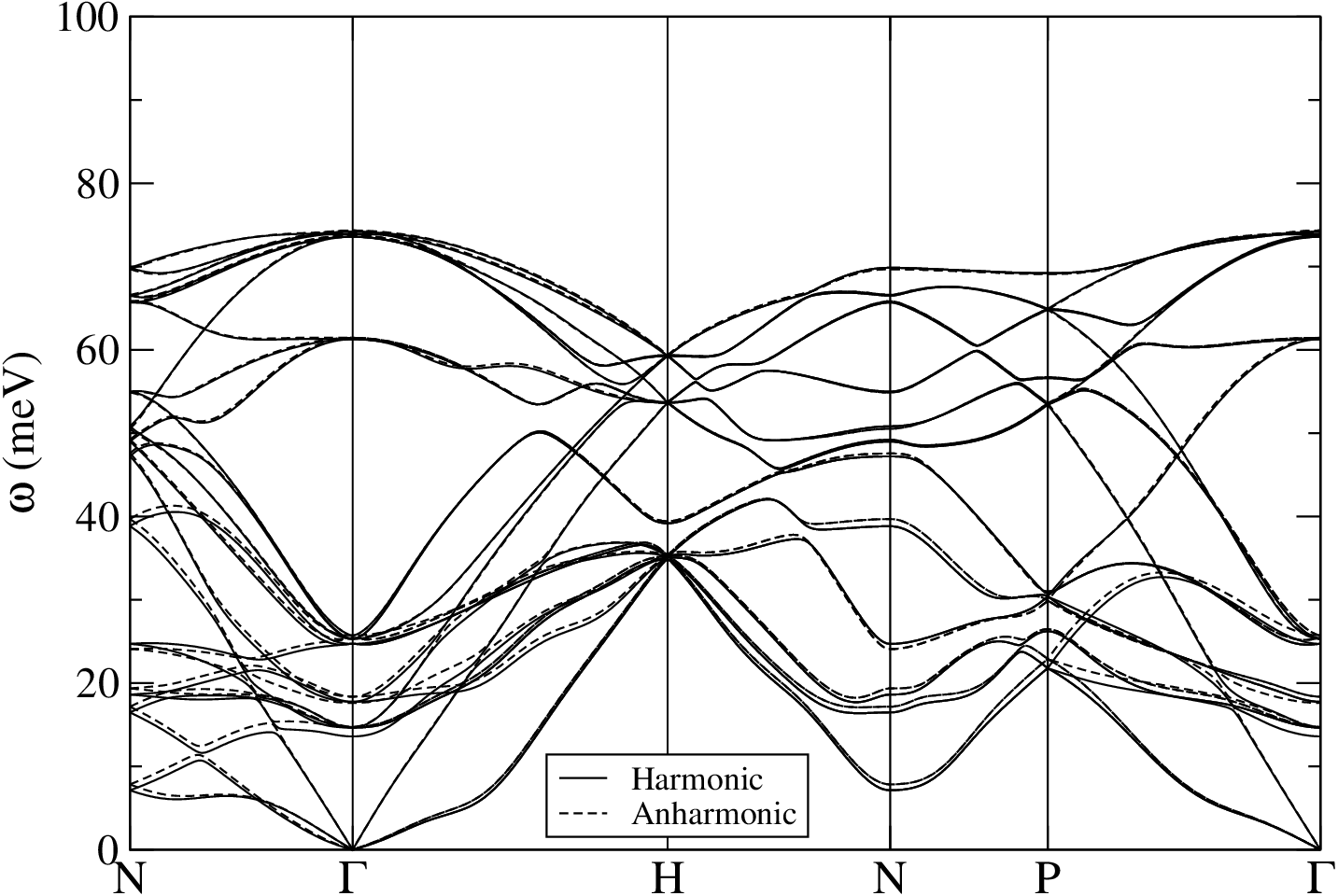}\label{ci16spectra15}}\hfill
\subfloat[(a)][19 GPa]{\includegraphics[width=0.48\linewidth]{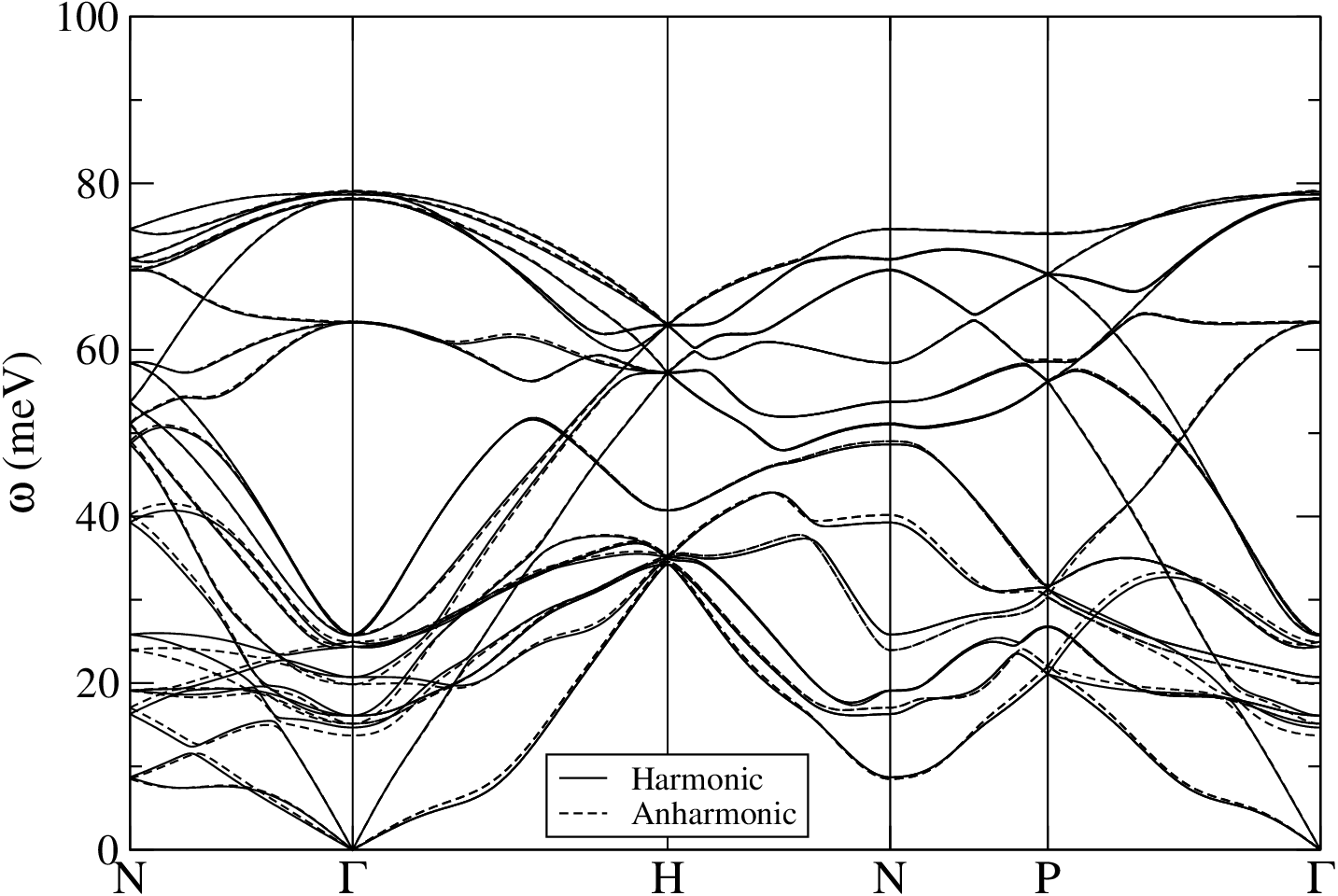}\label{ci16spectra18}}\\
\subfloat[(a)][27 GPa]{\includegraphics[width=0.48\linewidth]{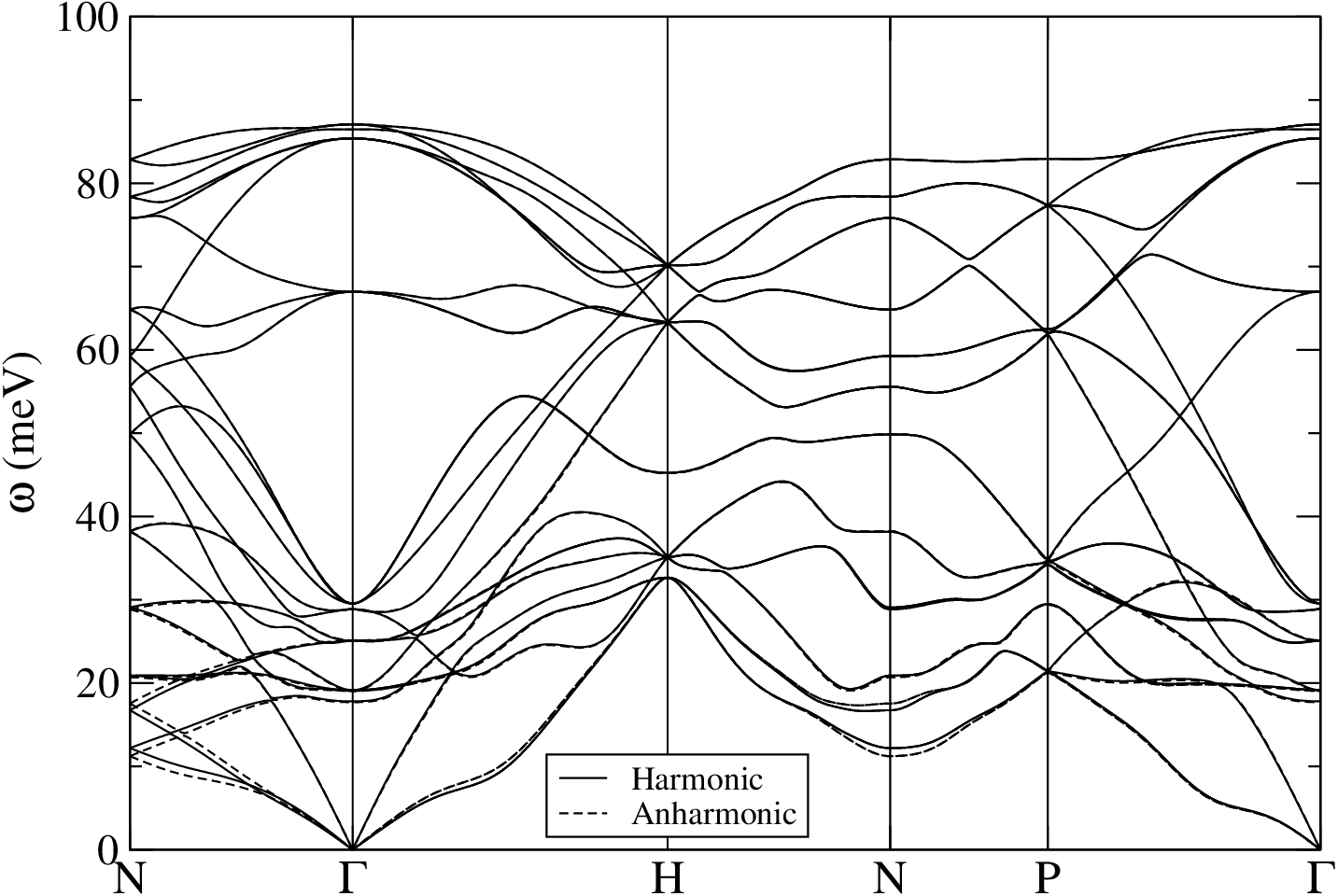}\label{ci16spectra27}}\hfill
\subfloat[(a)][44 GPa]{\includegraphics[width=0.48\linewidth]{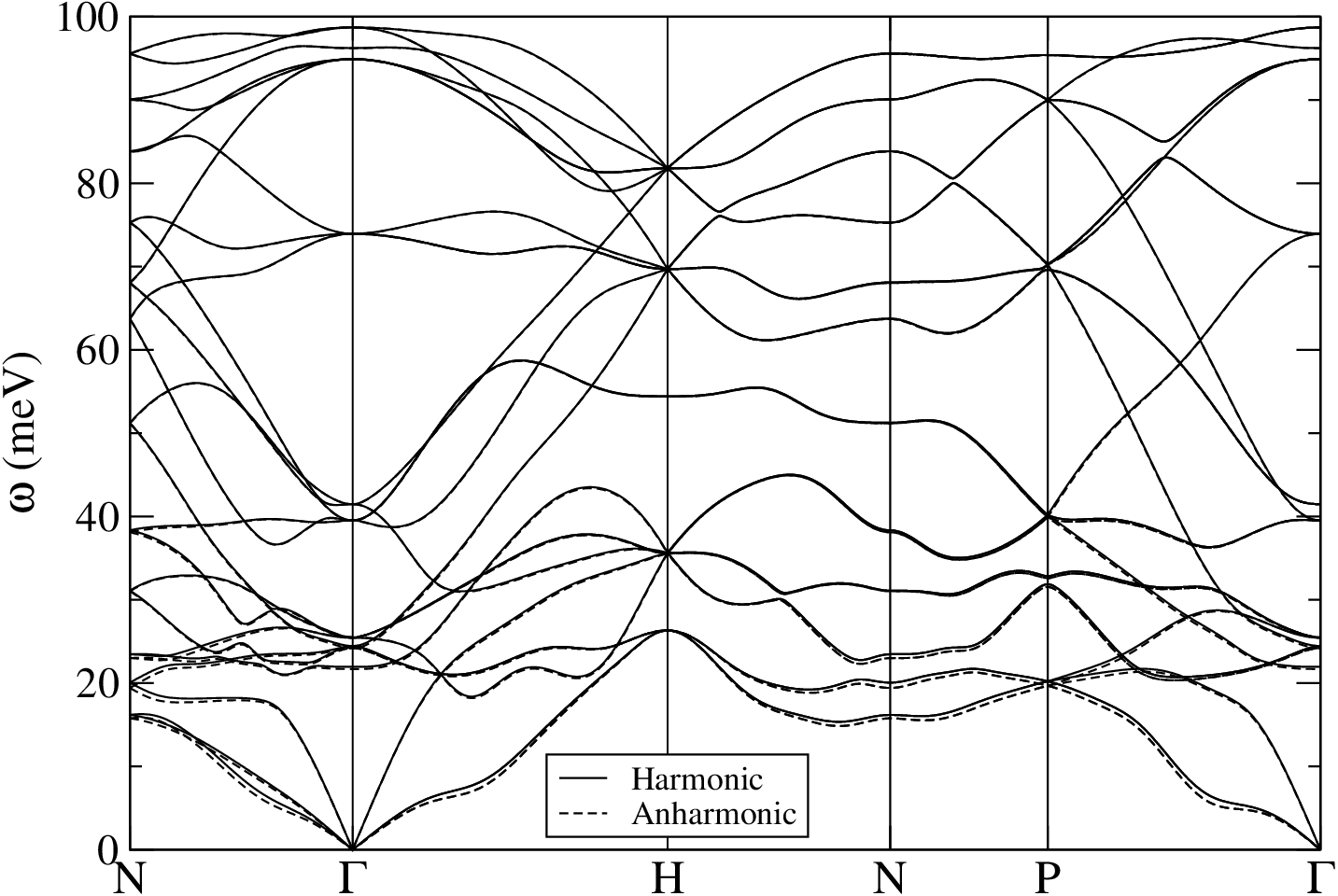}\label{ci16spectra44}}
\caption{Phonon spectra of fcc $\mathrm{^6Li}$ at different pressures.\label{ci16phononspectra}} 
\end{figure}

Electron-phonon matrix elements where calculated within DFPT, where converging the double Dirac delta in the equation for the phonon linewidth required a denser
$80\times80\times80$ \textbf{k}-point mesh. Superconducting $\mathrm{T_c}$ was calculated solving
isotropic Migdal-Eliashberg equations considering that for large electron-phonon coupling constants McMillan’s equation underestimates $\mathrm{T_c}$. 
In Fig. \ref{fcctcconvergence} we can see how converging $\mathrm{T_c}$ and $\lambda$ with the \textbf{q}-point grid becomes tedious due to the large contribution of $\mathrm{\mathbf{q}_{inst}}$ to the total electron-phonon coupling. Our chosen $9\times9\times9$ grid
overestimate $\mathrm{T_c}$ by 2.5 K comparing to the $12\times12\times12$ case while $\lambda$ is 0.2 larger. However, increasing the grid size would make the calculation really demanding and, our goal being to check whether anharmonicity could explain the 
anomalous isotope effect, this overestimation would only make anharmonic effects to be more visible. However, even in this case anharmonic effects are not big enough to explain the anomalous isotope effect. 
Moreover, we clearly see grids not containing $\mathrm{\mathbf{q}_{inst}}$ (dimensions not multiple of 3) yield smaller $\mathrm{T_c}$ and $\lambda$ values that the ones they do, and using such grids would obviously neglect how anharmonicity affects the 
electron-phonon coupling and superconductivity. 

\subsection{cI16 structure}

\begin{figure}[th]
\subfloat[(a)][Fcc  $\mathrm{^7Li}$ 26GPa]{\includegraphics[width=0.48\linewidth]{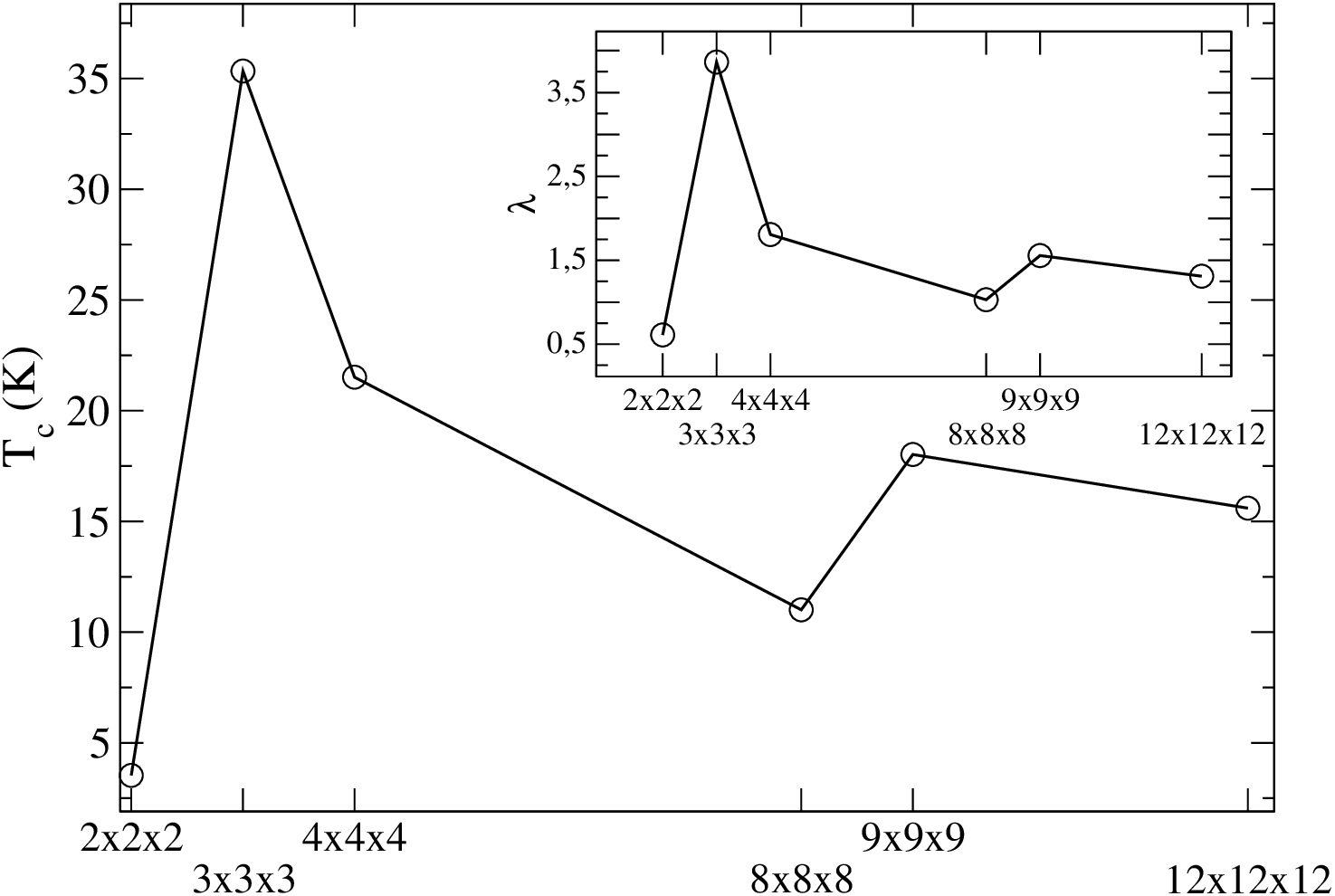} \label{fcctcconvergence}}\hfill
\subfloat[(b)][cI16 $\mathrm{^7Li}$ 27GPa]{\includegraphics[width=0.48\linewidth]{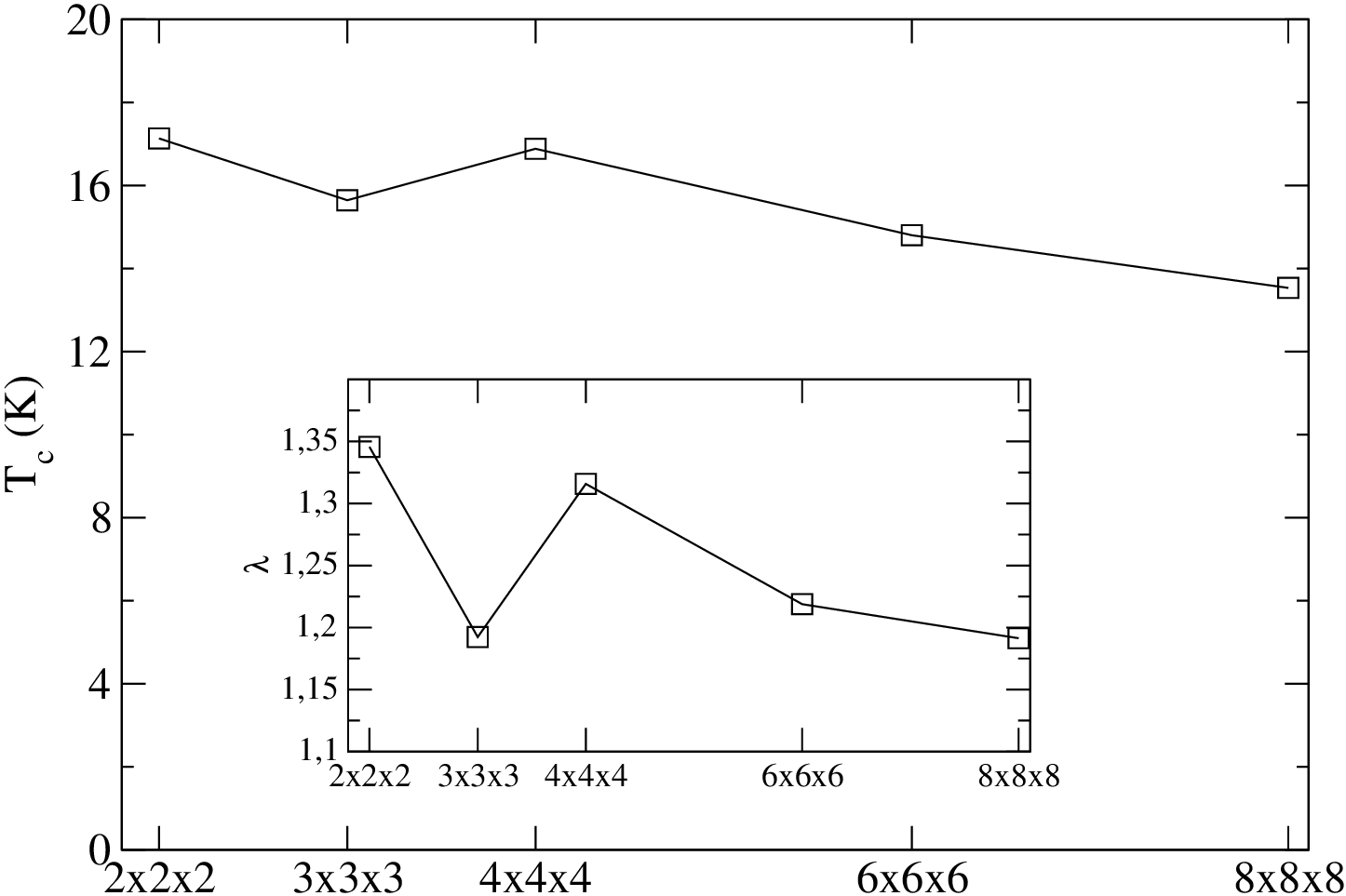}\label{ci16tcconvergence}}
\caption{Convergence of $\mathrm{T_c}$ and $\lambda$ with the \textbf{q}-grid for Fcc and cI16 $\mathrm{^7Li}$, using McMillan equation and $\mu^*=0.17$. \label{tcconvergence}} 
\end{figure}

Harmonic dynamical matrices have been obtained in a $6\times6\times6$ \textbf{q}-grid for every analyzed pressure and isotope. A proper convergence of phonon frequencies required a $16\times16\times16$ \textbf{k}-point grid and Methfessel-Paxton
smearing width of 0.01 Ry for electronic integrations in the first BZ. An energy cutoff of 65 Ry was necessary for expanding the wave-functions in the plane-wave basis. In Fig. \ref{ci16phononspectra} we show the phonon spectra obtained at 18, 27 and 44 GPa
for both $\mathrm{^6Li}$ and $\mathrm{^7Li}$ after Fourier interpolating the dynamical matrices from the $6\times6\times6$ grid to the desired path.  

Anharmonic dynamical matrices where obtained in a $2\times2\times2$ \textbf{q}-grid, commensurate to the supercell in which the SSCHA was perfomed. We interpolated the results to the finer $6\times6\times6$ grid with the same method as in the fcc case.
In this case anharmonicity has practically no influence on phonon frequencies at 27 and 44 GPa, while at 15 and 19 GPa low frequency modes are more noticeably affected.  
We can see this in Fig. \ref{ci16phononspectra} for $\mathrm{^6Li}$ (we do not show the result for $\mathrm{^7Li}$ as they are practically identical). 

Converging the double Dirac delta in the equation for the phonon linewidth required a $32\times32\times32$ \textbf{k}-point mesh. Superconducting $\mathrm{T_c}$ was calculated solving
isotropic Migdal-Eliashberg equations. 
Converging $\mathrm{T_c}$ within 1 K required to calculate the electron-phonon matrix elements in a $6\times6\times6$ \textbf{q}-point grid (see Fig. \ref{ci16tcconvergence}).

The cI16 structure (Space Group I-43d) has all the Li atoms placed in the Wyckoff 16c positions (conventional coordinates $(x,x,x)$ and all symmetry equivalent) , which has a free parameter $x$. 
As the SSCHA miniminization of the free energy is also performed with respect to $x$, final average atomic positions are
different from the harmonic or static ones. In principle, one should perform the electron-phonon coupling calculations in the new anharmonic atomic positions for each isotope and pressure. However, we checked that the impact on $\lambda$ and $T_c$ 
for $\mathrm{^6Li}$ at 19 GPa, where the change in $x$ is the greatest ($\Delta x=$0.004) is within the convergence criteria. 
Therefore, we use the electron-phonon coupling calculation calculated at the static equilibrium positions at each pressure for both isotopes.

\begin{figure}[t]
\subfloat[(a)][Fcc  $\mathrm{^7Li}$ 15GPa]{\includegraphics[width=0.48\linewidth]{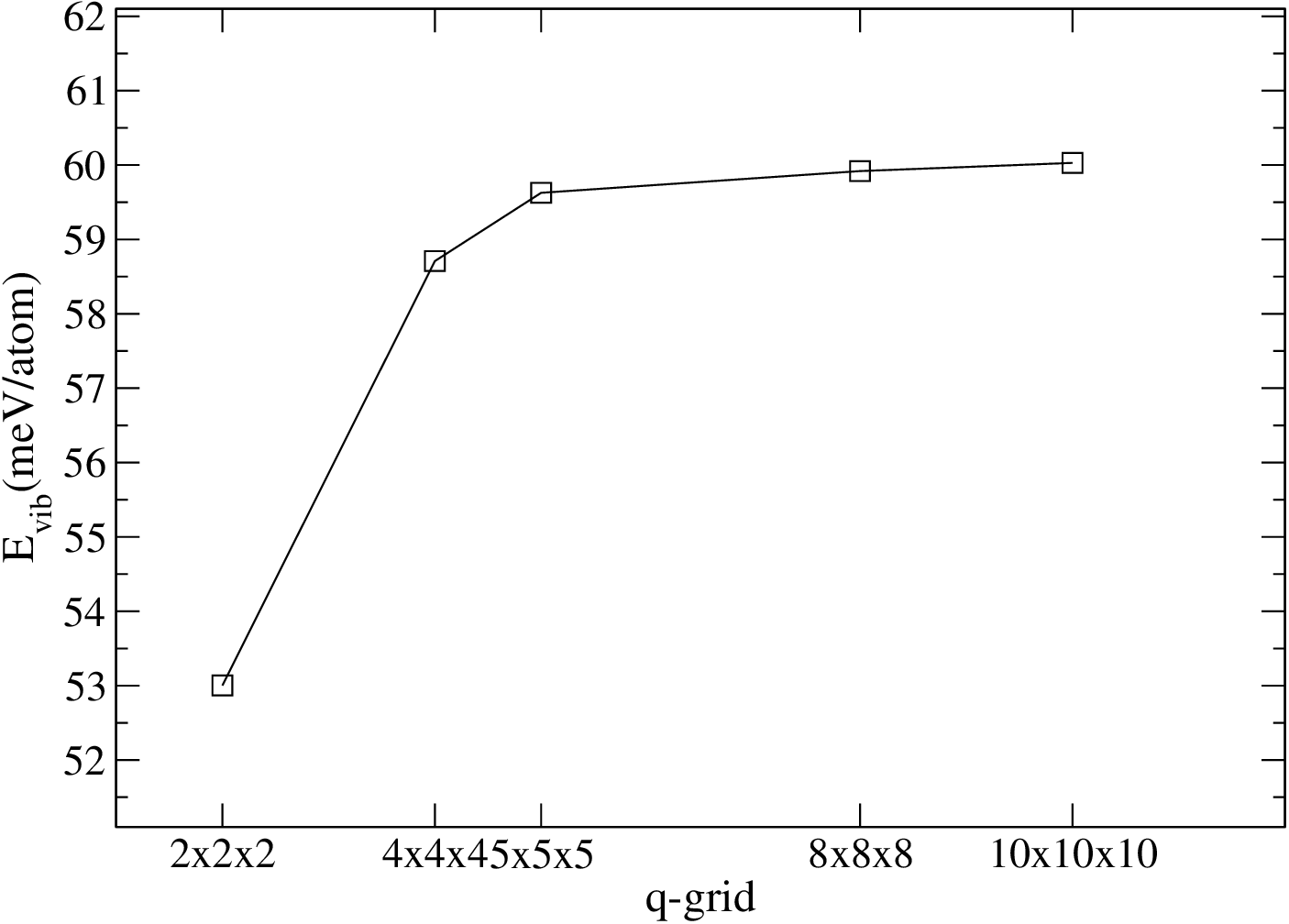} \label{fccevibconvergence}}\hfill
\subfloat[(b)][cI16 $\mathrm{^7Li}$ 27GPa]{\includegraphics[width=0.48\linewidth]{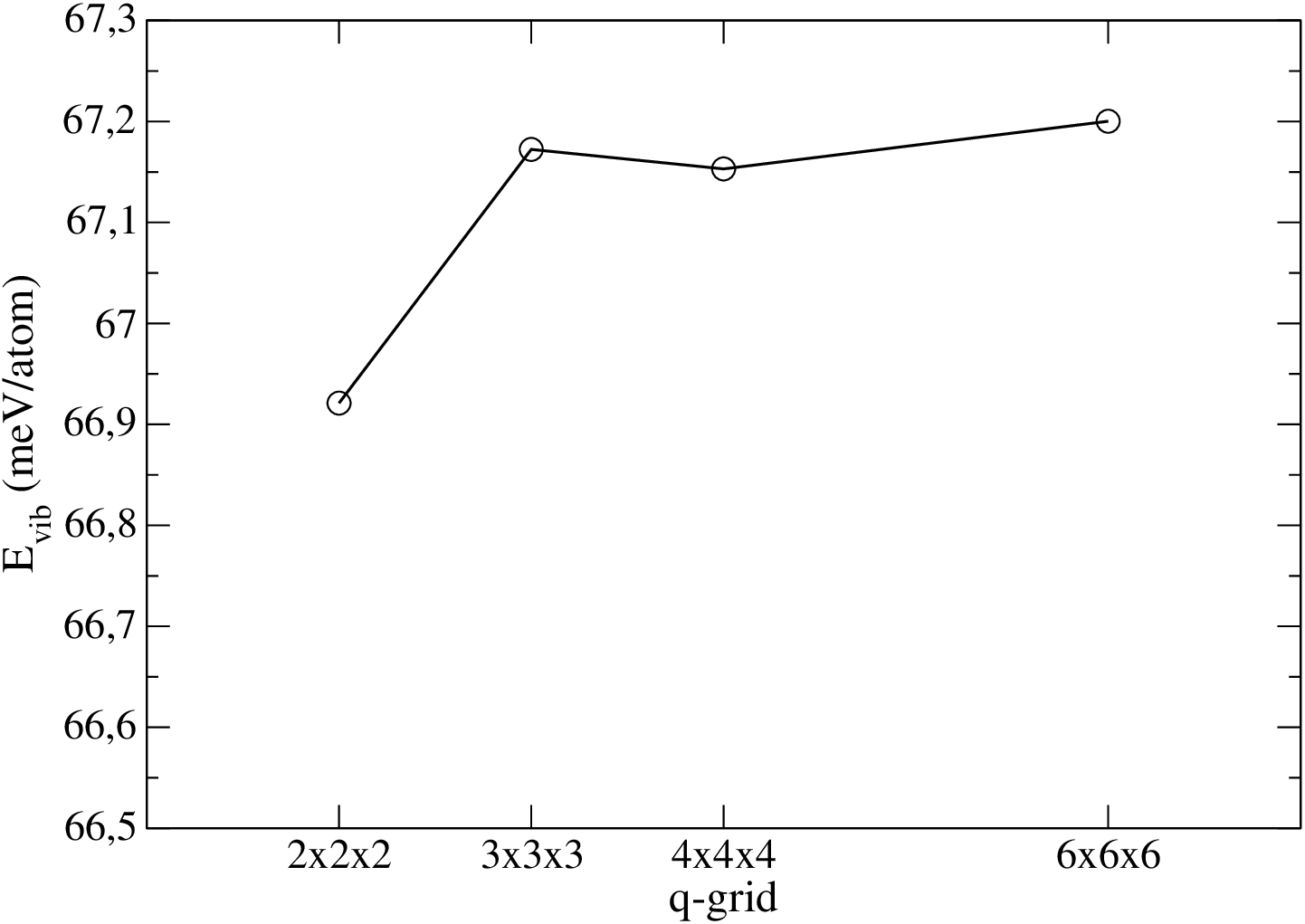}\label{ci16evibconvergence}}
\caption{Convergence of $E_{har}$ with the \textbf{q}-grid for Fcc and cI16 $\mathrm{^7Li}$\label{evibconvergence}} 
\end{figure}

\section{Enthalpy curves}

For obtaining the enthalpy $H=E_{T} + PV$ of the different structures, we calculated each contribution to the total energy $E_{T}=E_{el}+E_{v}$ at several unit-cell volumes and fitted them separately, 
due to the fact that the computational cost of a data point differs significantly from one contribution to another, as electronic energy $E_{el}$ is faster to compute than the vibrational $E_{v}$ one.

We calculated $E_{el}$ for fcc and ci16 Li for volumes per atom ranging from 50 to 100 $a_0^3$ with a step size of approximately 1.5 $a_0^3$. We fitted the data using a Birch-Murnaghan equation of state.
Due to the different properties of the phonon spectra, the vibrational contribution required a different treatment for each crystal structure. We find convenient to write the total vibrational contribution as
$E_v=E_{freq}+<V-\mathcal{V}>$\cite{PhysRevB.89.064302}, where $E_{freq}$ comes from the sum of the SSCHA frequencies over all the modes of the crystal and $<V-\mathcal{V}>$ comes from the difference of the actual anharmonic energy surface and the SSCHA
harmonic one. 
$E_{freq}$ can be further splitted into the harmonic contribution and the anharmonic correction, $E_{freq}=E_{har}+E_{anh}$, where $E_{har}$ is the energy coming from the harmonic frequencies.

For cI16 calculating harmonic dynamical matrices in a $2\times2\times2$ \textbf{q}-grid was enough to converge $E_{har}$ within 0.5 meV/atom. We calculated $E_{har}$ at seven different volumes, from 50 to 100 $a_0^3$, and fitted a
fourth order polynomial to the data points. We calculated $E_{anh}$ and $<V-\mathcal{V}>$ at four different volumes per atom (60,70,80 and 84 $a_0^3$) by perfoming SSCHA calculations in  $2\times2\times2$ supercells, and fitted the data with
a second order polynomial.


\begin{figure}[t]
 \includegraphics[width=0.7\linewidth]{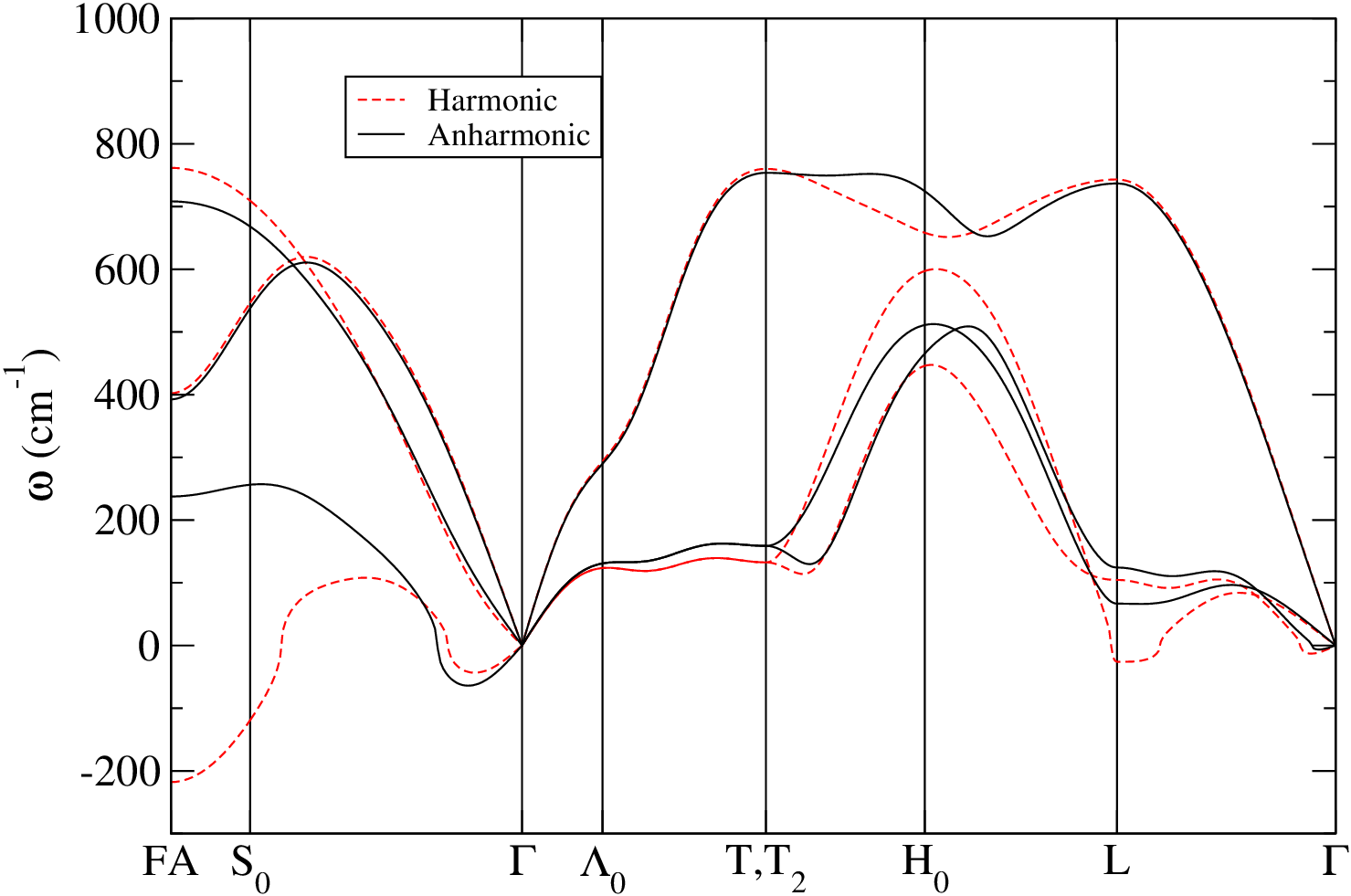}
 \caption{Phonon spectra of hR1 $\mathrm{^7Li}$ at around 40 GPa. Harmonic dynamical matrices have been explicitly calculated in a $6\times6\times6$ \textbf{q}-grid, while the anharmonic ones have been calculated in a $2\times2\times2$ grid
 and interpolated to the finer $6\times6\times6$ \textbf{q}-grid afterwards. The harmonic spectrum shows phonon instabilities in large regions of the BZ. Anharmonicity renormalizes strongly those instabilities, yielding real frequencies for every 
 \textbf{q}-point in the $2\times2\times2$ grid. However, after interpolating to the $6\times6\times6$ grid some modes remain unstable.   \label{hR1phonons}}
\end{figure}

Fcc Li presents a more complex situation due to the anomaly in the $\Gamma$-K path. We computed $E_{har}$ using a $8\times8\times8$ grid, which does not show any imaginary
frequency down to at least 65 $a_0^3$/atom (which corresponds to around 35 GPa), and converges $E_{har}$ within 0.2 meV/atom (see Fig. \ref{evibconvergence}). We calculated $E_{har}$ at seven different volumes, from 50 to 100 $a_0^3$, and fitted a
fourth order polynomial to the data points. 
To estimate the anharmonic contribution, we performed SSCHA calculations to obtain anharmonic dynamical matrices and $<V-\mathcal{V}>$ in a $3\times3\times3$ grid for four different volumes (66,72,77 and 84 $a_0^3$). 
To overcome the situation of using different grids for each contribution of the vibrational energy, we needed to treat $E_{anh}$ carefully. We interpolated our anharmonic dynamical matrices from the $3\times3\times3$ grid to a finer
$9\times9\times9$ one to obtain $E_{freq}$, and substracted the harmonic contribution in a $8\times8\times8$ grid as $E_{anh}=E_{freq}-E_{har}$, as imaginary frequencies prevent us obtaining $E_{har}$ in a $9\times9\times9$ grid. 
Finally, we fitted these four data points with a second order polynomial. 

%

\begin{figure}[t]
 \includegraphics[width=0.7\linewidth]{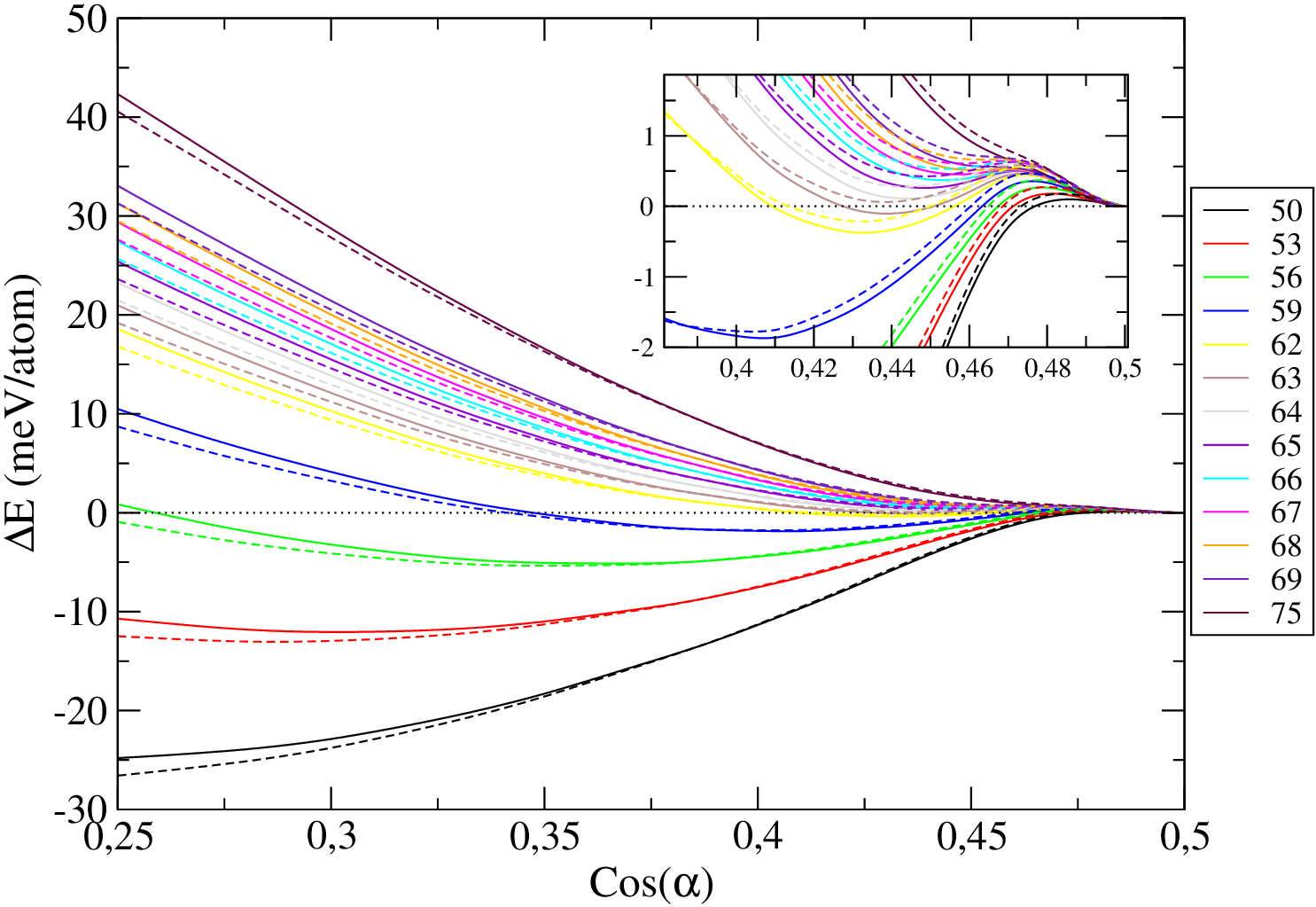}
 \caption{$\Delta E(\alpha,V)$ against $\cos{(\alpha)}$ for different unit-cell volumes (in atomic units). In the dashed curves vibrational energy is not included.  \label{hR1profiles}}
\end{figure}

For hR1 we have proceeded in a different way due to the fact that it shows plenty of imaginary frequencies in the harmonic phonon spectra (see Fig. \ref{hR1phonons}). 
These imaginary frequencies are strongly renormalized by anharmonicity and become real after applying the SSCHA in a $2\times2\times2$ \textbf{q}-grid. However, the interpolation method was not useful in this case as some of the interpolated
anharmonic matrices in a $6\times6\times6$ remained yielding imaginary frequencies. We overcame this situation making use of the similarity of hR1 with the fcc phase. 
If one chooses a rhombohedral unit cell, hR1 differs from fcc only by the rhombohedral angle $\alpha$. Thus, taking  $\alpha$ and the unit cell volume $V$ as variables, we can focus our attention to their associated potential energy surface. 
We define the total energy as $E_{T}(\alpha,V)=E_{T,fcc}(V)+\Delta E(\alpha,V)$, where $E_{T,fcc}(V)$ is the total energy of the fcc phase ($\alpha=$60º) and $\Delta E(\alpha,V)$  is the difference in energy due to the change in rhombohedral angle.
We only need to calculate $\Delta E(\alpha,V)$ in this case as we had previously calculated $E_{T,fcc}(V)$. $\Delta E(\alpha,V)$ is the sum of electronic and vibrational contributions. 
The electronic contribution $\Delta E_{el}(\alpha,V)$ is easily obtained by DFT total energy 
calculations. For obtaining the vibrational contribution  $\Delta E_{v}(\alpha)$  we assumed that it is independent of the unit cell volume. This way, we performed SSCHA calculations in $2\times2\times2$ supercells at four different $\cos(\alpha)$ values 
(0.25,0.35,0.412 and 0.5) for a single volume (60 $a_0^3$) and  fitted it with a $3^{rd}$ order polynomial.  
In Fig. \ref{hR1profiles} we show $\Delta E(V,\alpha)$ against $\alpha$ for different choices of the unit cell volume V, which is kept constant in each curve. 
Two relative minima can be distinguished below 70 $a_0^3$: one at $\cos\alpha=0.5$, which corresponds to the fcc structure, and another one corresponding to the hR1 phase, 
which even has a lower energy than the previous one for volumes smaller than 63 $a_0^3$. Plus, the angle at which this minimum occurs increases with decreasing volume. Above 70 $a_0^3$ hR1 could not exist as it lacks of a local energy minimum.

%

\begin{figure}[t]
 \includegraphics[width=0.7\linewidth]{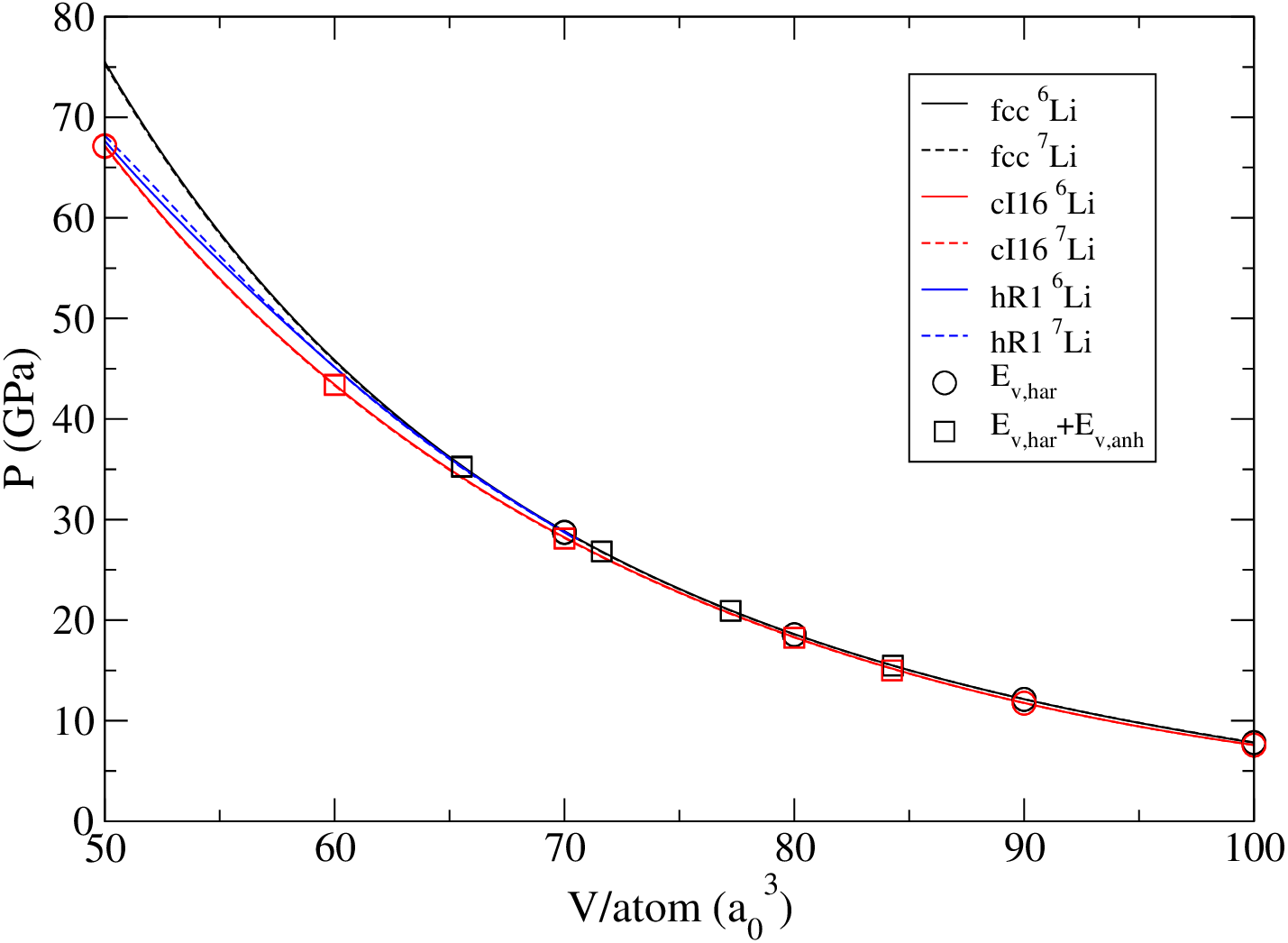}
 \caption{Pressure of fcc, cI16 and hR1 Li for both isotopes. The symbols show data points in which harmonic ($E_{har}$) and anharmonic ($E_{anh}$) vibrational data has been explicitly calculated. \label{pressure}}
\end{figure}

In Fig. \ref{pressure} we show the pressure vs. volume curves for each isotope and structure, obtained by taking the first derivative of $E_{T}$ with respect to the volume.

\bibliography{bibliografia}
\bibliographystyle{apsrev4-1}